\documentclass [12pt]{article}
\usepackage{amsfonts,amsmath,amssymb}
\usepackage{graphicx}
\usepackage{epic,eepic,color,graphicx}

\newfont{\twelvemsb}{msbm10 scaled\magstep1}
\newfont{\eightmsb}{msbm8}
\newfam\msbfam
\textfont\msbfam=\twelvemsb \scriptfont\msbfam=\eightmsb
\catcode`\@=11
\def\Bbb{\ifmmode\let\next\Bbb@\else
\def\next{\errmessage{Use \string\Bbb\space only in math mode}}\fi\next}
\def\Bbb@#1{{\fam\msbfam{{#1}}}}

\newcommand{\be}{\begin{equation}}
\newcommand{\ee}{\end{equation}}
\newcommand{\ba}{\begin{eqnarray}}
\newcommand{\ea}{\end{eqnarray}}
\newcommand{\nn}{\nonumber}

\begin{document}

\sloppy
\renewcommand{\thefootnote}{\fnsymbol{footnote}}
\newpage
\setcounter{page}{1} \vspace{0.7cm}
\begin{flushright}
01/03/10
\end{flushright}
\vspace*{1cm}
\begin{center}
{\bf The high spin expansion of twist sector dimensions:\\ the planar ${\cal N}=4$ super Yang-Mills theory}\\
\vspace{1.8cm} {\large Davide Fioravanti $^a$ and
Marco Rossi $^b$
\footnote{E-mail: fioravanti@bo.infn.it, rossi@cs.infn.it}}\\
\vspace{.5cm} $^a$ {\em Sezione INFN di Bologna, Dipartimento di Fisica, Universit\`a di Bologna, \\
Via Irnerio 46, Bologna, Italy} \\
\vspace{.3cm} $^b${\em Dipartimento di Fisica dell'Universit\`a
della Calabria and INFN, Gruppo collegato di Cosenza, I-87036
Arcavacata di Rende, Cosenza, Italy} \\
\end{center}
\renewcommand{\thefootnote}{\arabic{footnote}}
\setcounter{footnote}{0}
\begin{abstract}
{\noindent This review is devoted to collecting some results on the high spin expansion of (minimal) anomalous dimension. Thanks to the recent rationale on integrability,  planar ${\cal N}=4$ super Yang-Mills theory (or its $\text{AdS}_5\times\text{S}^5$ string counterpart) represents a very practicable field. Here the attention will be restricted to its sector of twist operators, although the analysis tools are quite general (in integrable theories). Some structures and ideas turn out to be general also for other sectors or gauge theories.}
\end{abstract}
\vspace{8cm}

\newpage

\section{Framework and beyond}
\setcounter{equation}{0}

We will move our investigation in within the maximally supersymmetric gauge theory in planar limit, {\it i.e.} for number of colours $N\rightarrow \infty$ and coupling $g_{YM}\rightarrow 0$,
so that the `t Hooft coupling
\begin{equation}
\lambda =  g_{YM }^2 N= 8 \pi^2 g^2 \,
\end{equation}
may stay finite. Among the different sectors (perturbatively closed under renormalisation), we also pick up the twist $sl(2)$ sector, spanned by local composite operators of trace form
\begin{equation}
{\mbox {Tr}} ({\cal D}^s {\cal Z}^L)+.... \, , \label {sl2op}
\end{equation}
where ${\cal D}$ is the (light-cone) covariant derivative acting in all the possible ways on the $L$ complex bosonic fields ${\cal Z}$. Trace ensures, of course, gauge invariance. The Lorentz spin of these operators is $s$ and $L$ is the $R$-charge which also coincides with the twist (classical dimension minus the spin). Besides, this sector may be described --
thanks to the AdS/CFT correspondence \cite{MWGKP} -- by spinning folded closed strings
on $\text{AdS}_5\times\text{S}^5$ spacetime with $\text{AdS}_5$ and $\text{S}^5$ angular momenta $s$ and $L$, respectively \cite{GKPII, FT}.

As being in a conformal model, suitable superpositions of operators form dilatation operator eigenvectors with definite dimensions (eigenvalues), which are made up of
a classical part plus an anomalous one. For instance, in the sector (\ref{sl2op})
this spectral problem shows up dimensions
\begin{equation}
\Delta(g,s,L) = L+s+\gamma (g,s,L) \, ,  \label{Delta}
\end{equation}
where $\gamma (g,s,L)$ is the anomalous part.
According to the AdS/CFT strong/weak coupling duality, the set of anomalous
dimensions of composite operators in ${\cal N}=4$ SYM coincides with
the energy spectrum of the $\text{AdS}_5\times\text{S}^5$ string theory (\cite{MWGKP,GKPII, FT} and references therein),
although the perturbative regimes are interchanged. The highly
nontrivial problem of evaluating the anomalous part in ${\cal
N}=4$ SYM was greatly simplified by the discovery of
integrability in the purely bosonic $so(6)$ sector at one loop \cite
{MZ}. Later on, this fact has been extended to all the gauge theory sectors and
at all loops in a way which shows up integrability in a weaker sense, but still furnishes the investigators many powerful tools \cite{BS}. More in detail, any
operator ({\it e.g.} of the form (\ref {sl2op})) has been thought of as a state of a 'spin chain', whose hamiltonian is, of course, the dilatation operator itself, although the latter does not have an explicit expression of the spin chain form, but for the first few loops. Nevertheless, the large size ({\it asymptotic}) spectrum has turned out to be exactly described by certain Asymptotic Bethe Ansatz-like equations (the so-called Beisert-Staudacher equations, cf. \cite {BS,BES} and references therein). In other words, the anomalous dimensions coincide with the energies given by the Bethe Ansatz solutions (or roots): this is, of course, a great simplification of the initial spectral problem.

Unfortunately, this works only for infinitely long operators: anomalous dimensions of operators with finite length depend not only on Asymptotic Bethe Ansatz (ABA) data but also on finite size 'wrapping' corrections: in the perturbative expansion wrapping effects are observed \cite {WRA} starting from the order $g^{2L}$. Recent progress has shown that a set of Thermodynamic Bethe Ansatz (TBA) equations \cite{TBA} or, together with certain additional information, an equivalent $Y$-system of functional equations \cite{Y} provides a basis for exact (any length at any coupling) predictions for anomalous dimensions of planar ${\cal N}=4$ SYM.

However, in the $sl(2)$ sector of ${\cal N}=4$ SYM relevance of wrapping effects seems to be reduced, even for short operators, if one goes to the high spin limit. For instance, findings of
\cite {BJL} have showed that, at least at twist two and up to five loops, wrapping corrections
start contributing at order $O\left ((\ln s )^2 /s^2 \right )$. This fact has pushed the idea of applying ABA techniques to the study of the high spin limit of twist operators, which - on the other hand - had already received much attention in the past literature.
Indeed, the high spin behaviour of the anomalous dimension
\begin{equation}
\gamma (g,s,L)=\Delta(g,s,L) -L-s \, , \label{gamma}
\end{equation}
shows a Sudakov behaviour
\begin{equation}
\gamma (g,s,L)=f(g)\ln s + \dots \, , \label{sud}
\end{equation}
determined by the so-called {\it universal} (since it does not depend on $L$ or the flavour) {\it scaling function}, $f(g)$ \cite{BGK, ES, FTT, BES, AM} . Actually, this behaviour is more general than in planar
${\cal N}=4$ SYM and the adjective {\it scaling} is due to the linear value of the {\it cusp anomalous dimension} with coefficient $f(g)/2$ while the cusp angle tends to infinity \cite{K} \footnote{Polyakov noticed as first that for cusped Wilson loop vacuum expectation value the charge renormalisation is not enough as in the non-cusped case, because of an extra logarithmic divergence due to the high bremsstrahlung at the cusp. He was led to consider cusps because of their importance in the loop dynamics (in euclidian space-time) \cite{Poly}.}. This large angle behaviour is due to the dominance by the lowest twist ($=2$) in the renormalisation of the vacuum expectation value of a cusped Wilson loop with a very large angle. In the end, for an infinite angle cusp ({\it i.e.} with one light-cone segment) $f(g)$ equals twice the cusp anomalous dimension of a light-cone Wilson loop \cite{KM}. Additionally, the one-loop problem (and thus $f(g)$) stays exactly the same for twist operators in QCD as long as the partonic helicities are aligned \cite{LIP,BDM}; this fact also has justified partially the great interest on the twist operators (\ref {sl2op}).

In general, the high spin limit of anomalous dimensions of twist $L$ operators goes on as a series of logarithmic (inverse)
powers\footnote {With $O\left ( (\ln s )^{-\infty} \right )$ we indicate terms going to zero faster than any inverse powers of $\ln s $.},
\ba
\gamma(g,s,L)= f(g) \ln s + f_{sl}(g,L) + \sum_{n=1}^\infty \gamma^{(n)}(g,L)  \, (\ln s)^{-n} + O\left ( (\ln s )^{-\infty} \right ) \, , \label {subl}
\ea
i.e. looks, at this order, like an expansion in the large 'size'  $\ln s$. Recently \cite {BES}, the leading term (i.e. $f(g)$) was obtained - in the hypothesis of being wrapping free - from the solution of a linear integral equation directly derived from the ABA via the root density approach. Moreover, $f(g)$ was carefully studied and tested both in the weak \cite {ES,BES} and strong coupling limit \cite {BBKS,CK,BKK,KSV}.

The sub-leading (constant) contribution $f_{sl}(g,L)$ received also much attention. In the ABA framework, it was shown \cite{FRS} to come from the solution of a non-linear integral equation (NLIE). Then, it was obtained \cite{BFR} starting from a linear integral equation (LIE). Explicit weak and strong coupling expansions performed using the LIE of \cite{BFR} are present in \cite {FZ,FGR4} and agree with some string theory computations \cite {BFTT}\footnote {Agreement between ABA \cite {GRO} and string theory \cite {BFTT} results is now established up to two loops.}. Importantly, after results of \cite {BJL} (cf. above) it may be inferred that both $f(g)$ and $f_{sl}(g,L)$ are exactly given by this approach based on the ABA (without wrapping corrections). Besides, both terms fix the $1/s$ coefficients via the reciprocity relation which may still be, consistently with \cite {BJL}, wrapping free.

The latter reasoning would support that even sub-logarithmic terms from ABA be exact, but it is even more unclear if and at what extent this might be trusted. In \cite {FGR5} we studied and computed in a systematic way the ABA contribution to them, by using a set of integral equations.
Possible wrapping corrections to our results are still to be determined. In this respect, in addition to TBA findings, also results on the string side of the correspondence in the spirit of \cite{BFTT} could be of fundamental importance. In fact, for instance, the one-loop contribution to the $1/\ln s$ term in the second of \cite{BFTT} comes partially from the $AdS_5$ modes (wrapping free) and partially from the $S^5$ ones (wrapping)\footnote{We wish to thank N. Gromov who pointed out this double origin as he sees it clearly from the algebraic curve approach which reproduces one-loop ABA plus some 'vacuum oscillations' (wrapping).}.  Nevertheless, a better understanding might come from exact (maybe TBA) calculations.

Wrapping effects should be negligible\footnote{This should be true at least for small values of $g$. But, again, this reasoning is not rigorous.} if one takes the infinite twist limit
\be
s\rightarrow \infty \, , \quad L\rightarrow \infty \, , \quad j=\frac {L-2}{\ln s}  \, \quad {\mbox {fixed}} \, ,
\label {jlimit}
\ee
and restricts to the study of the scaling functions $f_n(g)$, $f_n^{(r)}(g)$ appearing in the expansion
\be
\gamma (g,s,L)=\ln s \sum ^{\infty}_{n=0} f_n(g) j^n + \sum _{r
=0}^{\infty} (\ln s )^{-r}\sum
^{\infty}_{n=0}
f_n^{(r)}(g) j^n + O\left ( (\ln s )^{-\infty} \right ) \, . \label {gammaj}
\ee
For this reason, a reasonable amount of activity was also devoted to the study of limit
(\ref {jlimit}), using ABA techniques.
Results concerning $f_n(g)$ at weak coupling are present in \cite {FRS}. Strong
coupling behaviour of $f_n(g)$ was studied in \cite {FGR1,BK,FGR2,FGR3} by relying on linear integral equations. A detailed study of $f_n^{(r)}(g)$, $r\geq 0$, can be found in \cite {FIR,FGR5}.

In this review we want to give a summary of our activity in the framework of high spin limit
of twist operators. We will first give a general description of the method we have used and whose main tools are integral
equations derived from the
ABA equations. Then, we will briefly report the most important results we obtained.

This review is organised as follows.
In Section 2 we review the ABA equations for twist operators.
We illustrate the properties of the minimal anomalous dimension operators and
explain a technique which allows exact computations of ABA contributions to the anomalous
dimension by using the so-called Non-Linear Integral Equation (NLIE).
In Section 3 we specialise to the high spin limit and show that, if one neglects
$O\left ( (\ln s )^{-\infty} \right )$ terms, asymptotic anomalous dimension can be computed by relying on integral equations.
In Section 4 we study the high spin limit at fixed twist.
In Section 5 results in the scaling limit (\ref {jlimit}) are discussed.

\section{All-loop ABA and the (N)LIE}
\setcounter{equation}{0}

Let us recall the ABA equations \cite {BS,BES} for the $sl(2)$ sector,
\begin{equation}
\left ( \frac {u_k+\frac {i}{2}}{u_k-\frac {i}{2}} \right )^L \left
( \frac {1+\frac {g^2}{2{x_k^-}^2}}{1+\frac {g^2}{2{x_k^+}^2}}
\right )^L=\mathop{\prod^s_{j=1}}_{j\neq k}  \frac
{u_k-u_j-i}{u_k-u_j+i}  \left ( \frac {1-\frac
{g^2}{2x_k^+x_j^-}}{1-\frac {g^2}{2x_k^-x_j^+}} \right )^2
e^{2i\theta (u_k,u_j)}\, , \label {manyeq}
\end{equation}
where
\begin{equation}
x^{\pm}_k=x^{\pm}(u_k)=x(u_k\pm i/2) \, , \quad x(u)=\frac
{u}{2}\left [ 1+{ \sqrt {1-\frac {2g^2}{u^2}}} \right ] \, , \quad
\lambda =8\pi ^2 g^2 \, ,
\end{equation}
$\lambda $ being the 't Hooft coupling.
The so-called dressing factor \cite {AFS,BHL,BES} $\theta (u,v)$ is given by
\begin{equation}
\theta (u,v)=\sum _{r=2}^{\infty}\sum _{\nu =0}^{\infty} \beta
_{r,r+1+2\nu}(g)
[q_r(u)q_{r+1+2\nu}(v)-q_r(v)q_{r+1+2\nu}(u)] \, ,
\end{equation}
the functions $\beta _{r,r+1+2\nu}(g)=g^{2r+2\nu-2}2^{1-r-\nu}c_{r,r+1+2\nu}(g)$ being
\begin{eqnarray}
\beta _{r,r+1+2\nu}(g)&=&2 \sum _{\mu =\nu}^{\infty} \frac {g^{2r+2\nu+2\mu}}{2^{r+\mu+\nu}} (-1)^{r+\mu+1}\frac {(r-1)(r+2\nu)}{2\mu +1} \cdot \nonumber \\
&\cdot & \left ( \begin{array}{cc} 2\mu +1 \\ \mu -r-\nu+1
\end{array} \right )\left ( \begin{array}{cc} 2\mu +1 \\ \mu -\nu
\end{array} \right )              \zeta (2\mu +1)
\end{eqnarray}
and $q_r(u)$ being the density of the $r$-th charge
\begin{equation}
q_r(u)=\frac {i}{r-1} \left [ \left (\frac {1}{x^+(u)}\right
)^{r-1}-\left (\frac {1}{x^-(u)}\right )^{r-1} \right ] \, .
\end{equation}
It is now clear that
configurations of Bethe roots, i.e. solutions of (\ref {manyeq}),
and the corresponding eigenvalues of the energy are related
respectively to composite operators and their anomalous dimensions
in the $sl(2)$ sector of ${\cal N}=4$ SYM.

In the $sl(2)$ sector states of twist $L$ are
described by an even number $s$ of real Bethe roots $u_k$ which satisfy (\ref  {manyeq}).
Bethe roots localize in an interval $[-b,b]$
of the real line. In addition to Bethe roots, also $L$ real 'holes' \cite{BGK, BKP, BES, FRS, BFR} are present
(a better explanation of the nature of holes will be given in the following).
For any state, two holes reside
outside the interval $[-b,b]$ and the remaining $L-2$ holes lie
inside this interval. We will indicate with
$u_h^{(i)}$ these 'internal' holes.

In this paper we will focus on the minimal anomalous dimension state.
For such a state the positions of both roots and holes are symmetric with respect to the origin.
For what concerns the internal holes, they all concentrate near the origin, with no roots lying in between.

In general, an efficient way to treat states described by solutions of a non-linear set of Bethe Ansatz equations consists in writing a non-linear integral equation, which is completely equivalent to them. The non-linear integral equation is satisfied by the counting function $Z(u)$, which in the case (\ref {manyeq})
reads as
\be
Z(u)=\Phi (u)-\sum _{k=1}^{s}
\phi (u,u_k) \, , \label {Z}
\ee
where
\ba
\Phi (u) &=&-2L \arctan 2u -iL \ln \left ( \frac {1+\frac {g^2}{2{x^-(u)}^2}}{1+\frac {g^2}{2{x^+(u)}^2}} \right )\, ,  \label {Phi} \\
\phi (u,v)&=& 2\arctan (u-v) -2i \left [ \ln \left ( \frac {1-\frac {g^2}{2x^+(u)x^-(v)} }{1-\frac {g^2}{2x^-(u)x^+(v)}} \right )+i\theta (u,v)\right] \, . \label {phi}
\ea
It follows from its definition that the counting function $Z(u)$, as a function of the real variable $u$, is a monotonously decreasing function and that
\be
\lim _{u\rightarrow \pm \infty} Z(u)=\mp \pi (L+s) \, .
\ee
Therefore, there are $L+s$ real points $\upsilon _k$ such that $e^{iZ(\upsilon _k)}=(-1)^{L+1}$.
It is a simple consequence of the definition of $Z(u)$ that $s$ of them coincide with the Bethe roots. The remaining $L$ points are called 'holes' and their role will be of fundamental importance in what follows.
As we anticipated before, for the minimal anomalous dimension state the internal holes concentrate near the origin, i.e. their positions $u_h^{(i)}$ are determined by the relations
\be
Z(u_h^{(i)})=\pi (2h+1-L) \, , \quad h=1, \ldots , L-2 \, . \label {holcond}
\ee
Excited states are obtained by making different choices for the 'quantum numbers' $h$.
It follows from the structure of (\ref {holcond}) that, for any state, $u_h^{(i)}$ depend in a non-linear way on the counting function $Z(u)$.

The non-linear integral equation for $Z(u)$ is written by using a modification of the standard ideas underlying the procedure concerning the excited state NLIE  \cite {FMQR}, this modification being dictated by the physical situation with two 'important' external holes. Suppose that in the interval $[-b,b]$ of the real line $s$ Bethe roots and
$L-2$ holes are present. Then, by using Cauchy theorem we can express a sum
over the Bethe roots of an observable $O(u)$ as
\begin{eqnarray}
\sum _{k=1}^{s} O(u_k)&=&-\int _{-b}^{b} \frac {dv}{2\pi} O(v)
Z^{\prime}(v)+ {\mbox {Im}} \int _{-b}^{b}\frac {dv}{\pi}
O(v-i\epsilon)\frac {d}{dv} \ln
[1+(-1)^{L} \, e^{iZ(v-i\epsilon)}] + \nonumber  \\
&+& {\mbox {Im}} \int _{0}^{-\epsilon }\frac {dy}{\pi}
O(-b+iy)\frac {d}{dy} \ln [1+(-1)^{L} \, e^{iZ(-b+iy)}]+ \label {int} \\
&+& {\mbox
{Im}} \int _{-\epsilon }^{0}\frac {dy}{\pi} O(b+iy)\frac {d}{dy}
\ln [1+(-1)^{L} \, e^{iZ(b+iy)}]  - \sum _{h=1}^{L-2} O(u_h^{(i)}) \, . \nonumber
\end{eqnarray}
The right hand side of (\ref {int}) does not depend on $\epsilon >
0 $ as far as no poles of the integrands $O(w) \frac {d}{dw}\ln [1+
(-1)^{L}e^{iZ(w)}]$ lie in the region $|{\mbox {Im}}\, w |\leq \epsilon$,
$|{\mbox {Re}}\, w| \leq b$. In addition, $\epsilon$ must be kept sufficiently small,
in such a way that
\begin{equation}
\left | e^{iZ(z)}\right | < 1 \, , \label {appr}
\end{equation}
where $z$ belongs to the integration contour of the last three integral terms in (\ref {int}).
This condition is assured by the monotonicity of the counting function,
i.e. in our case $Z'(v)<0$, $v\in [-b,b]$.
We apply (\ref {int}) to the sum over the Bethe roots contained in (\ref {Z}):
\ba
Z(u)&=&\Phi (u) + \int _{-b}^{b} \frac {dv}{2\pi} \phi (u,v)
Z^{\prime}(v) + \sum _{h=1}^{L-2} \phi (u,u_h^{(i)}) -
 {\mbox {Im}} \int _{-b}^{b}\frac {dv}{\pi}
\phi (u,v-i\epsilon)\frac {d}{dv} \ln
[1+(-1)^{L} \, e^{iZ(v-i\epsilon)}] - \nn \\
\label {Zgen} \\
&-& {\mbox {Im}} \int _{0}^{-\epsilon }\frac {dy}{\pi}
\phi (u,-b+iy)\frac {d}{dy} \ln [1+(-1)^{L} \, e^{iZ(-b+iy)}]- {\mbox
{Im}} \int _{-\epsilon }^{0}\frac {dy}{\pi} \phi(u,b+iy)\frac {d}{dy}
\ln [1+(-1)^{L} \, e^{iZ(b+iy)}] \, . \nonumber
\ea
What we have obtained is a non-linear integral equation - for the counting function
$Z(u)$ - which describes - in a way which is alternative to the Bethe Ansatz equations -
the minimal anomalous dimension state.
Since Bethe roots are localised in an interval of the real axis
(i.e. $b<\ + \infty$), NLIE (\ref {Zgen}) is different from the equation
introduced in \cite {FMQR}, which contains integrations over the entire real axis.
This difference reveals crucial in the specific problem of twist operators in the $sl(2)$ sector.
Indeed, important simplifications to the structure of
(\ref {Zgen}) arise in the high spin limit: if we decide to neglect terms going to zero faster
than any inverse power of $\ln s$, then it is possible to
get rid of the last three non-linear terms in (\ref {Zgen}).
We are now going to show this important property.

Coming back to (\ref {int}), we first use (\ref {appr}) in order to replace all the $\ln [1+(-1)^{L}
e^{iZ(z)}]$ with $\sum \limits _{n=1}^{\infty}(-1)^{n+1}(-1)^{nL}
\frac {e^{inZ(z)}}{n}$.
We obtain:
\begin{eqnarray}
\sum _{k=1}^{s} O(u_k)&=&-\int _{-b}^{b} \frac {dv}{2\pi} O(v)
Z^{\prime}(v)+ {\mbox {Im}} \int _{-b}^{b}\frac
{dv}{\pi} O(v-i\epsilon)\frac {d}{dv} \sum _{n=1}^{\infty}(-1)^{n+1}(-1)^{nL}
\frac {e^{inZ(v-i\epsilon)}}{n} + \nonumber \\
&+& {\mbox {Im}} \int _{0}^{-\epsilon }\frac {dy}{\pi}
O(-b+iy)\frac {d}{dy} \sum _{n=1}^{\infty}(-1)^{n+1}(-1)^{nL} \frac {e^{inZ(-b+iy)}}{n} + \label{sumh} \\
&+& {\mbox
{Im}} \int _{-\epsilon }^{0}\frac {dy}{\pi} O(b+iy)\frac {d}{dy}
\sum _{n=1}^{\infty}(-1)^{n+1}(-1)^{nL}\frac {e^{inZ(b+iy)}}{n} -  \sum _{h=1}^{L-2} O(u_h^{(i)}) \, . \nonumber
\end{eqnarray}
In order to evaluate the non-linear terms in this expression,
\begin{eqnarray}
NL&=&{\mbox {Im}} \int _{-b}^{b}\frac {dv}{\pi} O(v-i\epsilon)\frac
{d}{dv} \sum _{n=1}^{\infty}(-1)^{n+1}(-1)^{nL}\frac {e^{inZ(v-i\epsilon)}}{n} + \nonumber \\
&+& {\mbox {Im}} \int _{0}^{-\epsilon }\frac {dy}{\pi}
O(-b+iy)\frac {d}{dy} \sum _{n=1}^{\infty}(-1)^{n+1}(-1)^{nL} \frac {e^{inZ(-b+iy)}}{n} + \nonumber \\
&+& {\mbox {Im}} \int _{-\epsilon }^{0}\frac {dy}{\pi} O(b+iy)\frac {d}{dy}
\sum _{n=1}^{\infty}(-1)^{n+1}(-1)^{nL} \frac {e^{inZ(b+iy)}}{n} \, , \nonumber
\end{eqnarray}
we first assume that it is possible to exchange
the series with the integrals. Then, we use the following
formul{\ae}
\begin{eqnarray}
&&\int ^ v dv O(v-i\epsilon) \frac {d}{dv}e^{inZ(v-i\epsilon)}=\left . e^{inx} \sum _{k=0}^{\infty}
\left (\frac {i}{n}\right )^k \frac {d^k}{dx^k} O[Z^{-1}(x)] \right |_{x=Z(v-i\epsilon)} \, , \label {fo1}  \\
&&\int ^ y dy O(\pm b+iy) \frac {d}{dy}e^{inZ(\pm b+iy)}=\left. e^{inx} \sum _{k=0}^{\infty}
\left (\frac {i}{n}\right )^k \frac {d^k}{dx^k} O[Z^{-1}(x)]\right |_{x=Z(\pm b+iy)} \, . \label {fo2}
\end{eqnarray}
We need to remark that results (\ref {fo1}, \ref {fo2}) are correct if the above infinite sums make sense
either as convergent or asymptotic series.
In our case, we can use (\ref {fo1}, \ref {fo2}), since the series we will get are asymptotic.

Using (\ref {fo1}, \ref {fo2}) in the evaluation of the above non-linear terms, the dependence on $\epsilon $ cancels out, as it should be, and we are left with
\begin{equation}
NL = \sum _{n=1}^{\infty}\frac {(-1)^{n+1}(-1)^{nL}}{\pi n } \left
. \Bigl [ \sum _{k=0}^{\infty} \frac {i^{2k} }{n^{2k} }\sin n x
\frac {d^{2k}} {dx^{2k}} O(Z^{-1}(x)) + \sum _{k=0}^{\infty} \frac
{i^{2k} }{n^{2k+1} } \cos n x \frac {d^{2k+1}} {dx^{2k+1}}
O(Z^{-1}(x))
 \Bigr ]\right |_{x=Z(-b)}^{x=Z(b)} \, . \label {NLseries}
\end{equation}
Now, we are allowed to choose $b$ in such a way that $e^{iZ(\pm b)}=(-1)^{L} $:
therefore, since $O(v)$ is bounded, all the terms in (\ref
{NLseries}) proportional to the sine function are zero.
Thus, we are left only with terms containing the
cosine function, i.e., after summing over $n$,
\begin{equation}
NL=-\sum _{k=0}^{\infty}\frac {(2\pi)^{2k+1} }{(2k+2)!}
B_{2k+2}\left . \left ( \frac {1}{2} \right) \Bigl [ \frac
{\partial} {\partial  x^{2k+1}} O(Z^{-1}(x))
 \Bigr ]\right |_{x=Z(-b)}^{x=Z(b)} \, , \label {NLseries2}
\end{equation}
where $B_k(x)$ is the Bernoulli polynomial.
Relation (\ref {NLseries2}) allows to write the final formula:
\be
\sum _{k=1}^{s} O(u_k)=-\int _{-b}^{b} \frac {dv}{2\pi} O(v)
Z^{\prime}(v)-\sum _{h=1}^{L-2} O(u_h^{(i)})-\sum _{k=0}^{\infty}\frac {(2\pi)^{2k+1} }{(2k+2)!}
B_{2k+2}\left . \left ( \frac {1}{2} \right) \Bigl [ \frac
{\partial} {\partial  x^{2k+1}} O(Z^{-1}(x))
 \Bigr ]\right |_{x=Z(-b)}^{x=Z(b)} \, . \label {sumexpr}
\ee
The following expressions
\ba
\left . \frac {d} {dx } O(Z^{-1}(x)) \right |_{x=Z(-b)}^{x=Z(b)}&=& 2\frac{O'(b)}{Z'(b)} \, , \nn \\
\left . \frac {d^2}{dx^2} O(Z^{-1}(x)) \right |_{x=Z(-b)}^{x=Z(b)} &=& 2\frac {O''(b)-O'(b)\frac {Z''(b)}{Z'(b)}}
{[Z'(b)]^2} \, , \nn \\
\left . \frac {d^3} {d  x ^3} O(Z^{-1}(x)) \right |_{x=Z(-b)}^{x=Z(b)}&=& 2\frac{O^{'''}(b)-3O''(b)\frac{Z''(b)}{Z'(b)}-O'(b)\frac{Z'''(b)}{Z'(b)}+3O'(b)\frac{(Z''(b))^2}{(Z'(b))^2}}{[Z'(b)]^3}
\nn
\ea
give an idea of the form of the first nonlinear terms appearing in (\ref {sumexpr}).
Next step is to apply (\ref {NLseries2}) to the non-linear integral terms (the last three ones)
contained in (\ref {Zgen}). We have to replace $O(v)$ with $-\phi (u,v)$. From the form of
(\ref {NLseries2}) we realise that such non-linear terms are proportional to derivatives
\be
\left. \frac{d^n}{dv^n}\phi (u,v) \right |_{\pm b} \, , \quad n \geq 1 \, .
\ee
When $s \rightarrow \infty$, $b = \frac{s}{2}\left (1+O\left ( \frac{1}{s}\right) \right )$ and for such derivatives
\be
\left. \frac{d^n}{dv^n}\phi (u,v) \right |_{\pm b} = O\left (\frac{1}{b^{n+1}}\right)  \, .
\ee
On the other hand, for the derivatives of the counting function we can give the estimates
\be
\left. \frac{d^n}{dv^n}Z(v) \right |_{\pm b} = O\left ( \frac{1}{b^{n}}\right ) \, .\label {Zpmb}
\ee
Putting all together, we conclude that when the spin $s\rightarrow + \infty$ the non-linear integral terms contained in (\ref {Zgen}) are $ O(1/b)=O\left ( (\ln s )^{-\infty} \right )$.
Therefore, if we decide to neglect terms $O\left ( (\ln s )^{-\infty} \right )$ and to focus only on corrections $O((\ln s)^{-n})$, we are entitled not to consider all the non-linear integral terms in (\ref {Zgen}).
In this case we are left with
\be
Z(u)=\Phi (u) + \int _{-b}^{b} \frac {dv}{2\pi} \phi (u,v)
Z^{\prime}(v) + \sum _{h=1}^{L-2} \phi (u,u_h^{(i)}) + O\left ( (\ln s )^{-\infty} \right ) \, , \label {Zhigh}
\ee
and this equation has to be solved together with condition (\ref {holcond}), which fixes the holes positions $u_h^{(i)}$
in terms of the counting function $Z(u)$. It is important to remark that within our approximations non-linearity with respect to $Z(u)$ enters only through (\ref {holcond}).

Equations (\ref {Zhigh}, \ref {holcond}) are our starting point for the study of the high spin limit.
They will be worked out in next section.

\section{Integral equations for the logarithmic terms}
\setcounter{equation}{0}

Let us consider integral equation (\ref {Zhigh}) satisfied by the counting function in the high spin limit.
Passing to derivatives, we define
\be
\sigma (u)=Z'(u) \, .
\ee
We have
\be
\sigma (u)= \Phi '(u) + \int _{-b}^{b} \frac {dv}{2\pi} \frac{d}{du}\phi (u,v)
\sigma (v) + \sum _{h=1}^{L-2} \frac{d}{du} \phi (u,u_h^{(i)}) + O\left ( (\ln s )^{-\infty} \right ) \, .
\ee
We now decompose such equation in its one loop (with index $0$) and higher than one loop (with index $H$) contributions.
We set $\Phi (u)=\Phi _0 (u) + \Phi _H(u)$, $\phi (u,v)=\phi _0(u,v)+\phi _H(u,v)$,
$\sigma (u)=\sigma _0(u)+\sigma _H(u)$, where
\ba
\Phi _0(u) &=&-2L \arctan 2u \, , \quad \Phi _H(u)=-iL \ln \left ( \frac {1+\frac {g^2}{2{x^-(u)}^2}}{1+\frac {g^2}{2{x^+(u)}^2}} \right )\, ,  \label {Phidec} \\
\phi _0(u,v)&=& 2\arctan (u-v) \, , \quad \phi _H(u,v)=-2i \left [ \ln \left ( \frac {1-\frac {g^2}{2x^+(u)x^-(v)} }{1-\frac {g^2}{2x^-(u)x^+(v)}} \right )+i\theta (u,v)\right] \, , \label {phidec}
\ea
and where, after neglecting quantities which are $O\left ( (\ln s )^{-\infty} \right )$,
$\sigma _0 (u)$ and $\sigma _H(u)$ satisfy
\ba
\sigma _0(u)&=& \Phi _0'(u) + \int _{-b_0}^{b_0} \frac {dv}{2\pi} \frac{d}{du}\phi _0(u,v)
\sigma _0(v) + \sum _{h=1}^{L-2} \frac{d}{du} \phi _0(u,\bar u_h^{(i)}) + O\left ( (\ln s )^{-\infty} \right ) \, , \label {1loop} \\
\sigma _H(u)&=&
 \Phi _H'(u) + \int _{-b}^{b} \frac {dv}{2\pi} \frac{d}{du}\phi _H(u,v)
\sigma _H (v) + \int _{-b_0}^{b_0} \frac {dv}{2\pi} \frac{d}{du}\phi _H(u,v)
\sigma _0 (v) + \nonumber \\
&+& \int _{-b}^{b} \frac {dv}{2\pi}\frac{d}{du}\phi _0(u,v)\sigma _H(v) +\sum _{h=1}^{L-2} \frac{d}{du} [\phi (u,u_h^{(i)}) - \phi _0 (u,\bar u_h^{(i)})]+ O\left ( (\ln s )^{-\infty} \right ) \, . \label {higlps}
\ea
In (\ref {1loop}, \ref {higlps}) the notation $\bar u_h^{(i)}$ stands for the one loop component of the
position of the $h$-th internal hole.
In order to solve the one loop equation (\ref {1loop}), we consider (see 3.52 of \cite {BFR}) a function
$\sigma _0 ^{(s)}(u)$, whose Fourier transform\footnote {We adopt the following definition for the Fourier
transform $\hat f(k)$ of a function $f(u)$: \be \hat f(k)=\int _{-\infty}^{+\infty} du e^{-iku}f(u) \, . \ee} reads
\be
\hat \sigma _0^{(s)}(k)= -4\pi \frac {\frac {L}{2}-e^{-\frac {|k|}{2}}\cos (k
s/\sqrt{2})} {2\sinh \frac {|k|}{2}}+2\pi \frac {e^{-\frac
{|k|}{2}}}{2\sinh \frac {|k|}{2}}\sum _{h=1}^{L-2}e^{ik\bar u_h^{(i)}}-(4\pi \ln 2) \delta (k)  \,
. \label {sigma0}
\ee
Function (\ref {sigma0}) satisfies the following important property \cite {BFR},
\be
\int _{-b_0}^{b_0} du f(u) \sigma _0 (u) = \int _{-\infty}^{+\infty} du f(u) \sigma _0 ^{(s)}(u) + O\left ( (\ln s )^{-\infty} \right ) \, ,
\ee
which allows to extend to the whole real axis the integrations involving $\sigma _0(u)$: a look at
many loops equation (\ref {higlps}) shows that this is just the property we need in order to try to solve it.

It follows from results in \cite {ES} and from numerical checks that
we can extend to the entire real axis the integrations involving $\sigma _H(u)$: what we are missing
are $O\left ( (\ln s )^{-\infty} \right )$ terms.
Putting all these pieces together, the equation satisfied by $\sigma _H(u)$ is
\ba
\sigma _H(u)&=&
 \Phi _H'(u) + \int _{-\infty}^{+\infty} \frac {dv}{2\pi} \frac{d}{du}\phi _H(u,v)
[\sigma _H (v) +\sigma _0^{(s)} (v)] + \nonumber \\
&+& \int _{-\infty}^{+\infty} \frac {dv}{2\pi}\frac{d}{du}\phi _0(u,v)\sigma _H(v) +\sum _{h=1}^{L-2} \frac{d}{du} [\phi (u,u_h^{(i)}) - \phi _0 (u,\bar u_h^{(i)})]+ O\left ( (\ln s )^{-\infty} \right ) \, . \label {higlps2}
\ea
It is now convenient to pass to Fourier transforms.
We have
\ba
\hat \Phi _H(k)&=& \frac{2\pi L}{ik} e ^{-\frac{|k|}{2}} [1-J_0(\sqrt{2}gk)] \, , \\
\hat \phi _H(k,t)&=& -8i \pi ^2 \frac{e^{-\frac {|t|+|k|}{2}}}{k|t|}\Bigl [ \sum _{r=1}^{\infty} r (-1)^{r+1}J_r({\sqrt
{2}}gk) J_r({\sqrt {2}}gt)\frac {1-{\mbox {sgn}}(kt)}{2}
 + \nonumber \\
&+&{\mbox {sgn}} (t) \sum _{r=2}^{\infty}\sum _{\nu =0}^{\infty}
c_{r,r+1+2\nu}(g)(-1)^{r+\nu} \Bigl (
J_{r-1}({\sqrt {2}}gk) J_{r+2\nu}({\sqrt {2}}gt)-  \nn  \\
&-& J_{r-1}({\sqrt {2}}gt) J_{r+2\nu}({\sqrt {2}}gk)\Bigr ) \Bigr ] \, . \label {sigmaeq2}
\ea
Therefore, the Fourier transform of (\ref {higlps2}) reads:
\ba
\hat \sigma _H(k) &=&\frac {\pi L}{\sinh \frac {|k|}{2}}[1-J_0({\sqrt {2}}gk)]
+  \frac {1}{\sinh \frac {|k|}{2}} \int _{-\infty }^{+\infty}
\frac {dt}{|t|} \Bigl [ \sum _{r=1}^{\infty}
r (-1)^{r+1}J_r({\sqrt {2}}gk) J_r({\sqrt {2}}gt)\frac {1-{\mbox {sgn}}(kt)}{2}
e^{-\frac {|t|}{2}} + \nonumber \\
&+&{\mbox {sgn}} (t) \sum _{r=2}^{\infty}\sum _{\nu =0}^{\infty} c_{r,r+1+2\nu}(g)
(-1)^{r+\nu}e^{-\frac {|t|}{2}} \Bigl (
J_{r-1}({\sqrt {2}}gk) J_{r+2\nu}({\sqrt {2}}gt)-
J_{r-1}({\sqrt {2}}gt)
J_{r+2\nu}({\sqrt {2}}gk)\Bigr ) \Bigr ] \cdot \nonumber \\
&\cdot& \Bigl [ \hat \sigma _H(t) +\hat \sigma _0^{(s)}(t)
+ 2\pi \sum _{h=1}^{L-2}e^{i t u_h^{(i)}} \Bigl ] + 2\pi  \frac {e^{-\frac {|k|}{2}}}{2 \sinh \frac {|k|}{2}}
\sum _{h=1}^{L-2} \left[ e^{-i k u_h^{(i)}}-e^{-i k \bar u_h^{(i)}} \right]+ O\left ( (\ln s )^{-\infty} \right )
\, . \label{Skeq}
\ea
Inserting in (\ref {Skeq}) the expression (\ref {sigma0}) for $\hat \sigma _0^{(s)}(t)$,
introducing the 'magic' kernel $\hat K(t,t')$, defined in \cite {BES} as
\be
\hat K(t,t')=\frac{2}{tt'}\left [ \sum _{n=1}^{\infty}n J_n (t) J_n (t') + 2 \sum _{k=1}^{\infty} \sum _{l=0}^{\infty} (-1) ^{k+l}c_{2k+1,2l+2}(g) J_{2k}(t) J_{2l+1}(t') \right ] \, , \label {magic}
\ee
and restricting to $k\geq 0$, we finally get the integral equation,
\begin{eqnarray}
\hat \sigma _H(k) &=&\frac {\pi L}{\sinh \frac {k}{2}}[1-J_0({\sqrt {2}}gk)]
- g^2 \frac {k}{\sinh \frac {k}{2}} \int _{0 }^{+\infty}
dt e^{-\frac{t}{2}} \hat K (\sqrt{2}gk,\sqrt{2}gt) \cdot \nonumber \\
&\cdot &
\Bigl \{ \hat \sigma _H(t)
-4\pi \frac {\frac {L}{2}-e^{-\frac {|t|}{2}}\cos (ts /{\sqrt 2}) }{2 \sinh \frac {|t|}{2}}
-(4\pi \ln2 )\delta (t)
+2\pi  \frac {e^{-\frac {|t|}{2}}}{2 \sinh \frac {|t|}{2}}  \sum _{h=1}^{L-2}\cos {t \bar u_h^{(i)}}+ 2\pi \sum
_{h=1}^{L-2}\cos{t u_h^{(i)}} \Bigl \}+ \nonumber \\
&+& \pi  \frac {e^{-\frac {k}{2}}}{ \sinh \frac {k}{2}}  \sum _{h=1}^{L-2} \left[ \cos {k u_h^{(i)}}-\cos {k \bar u_h^{(i)}} \right]
\, . \label{sigmakeq2}
\end{eqnarray}
Since we are interested in the computation of the anomalous dimension,
we go back to the general formula (\ref {sumexpr}) for the evaluation of observables,
\be
\sum _{k=1}^{s} O(u_k)=-\int _{-b}^{b} \frac {dv}{2\pi} O(v)
Z^{\prime}(v)-\sum _{h=1}^{L-2} O(u_h^{(i)})-\sum _{k=0}^{\infty}\frac {(2\pi)^{2k+1} }{(2k+2)!}
B_{2k+2}\left . \left ( \frac {1}{2} \right) \Bigl [ \frac
{\partial} {\partial  x^{2k+1}} O(Z^{-1}(x))
 \Bigr ]\right |_{x=Z(-b)}^{x=Z(b)} \, . \label {sumobs}
\ee
If we specialise to the energy (anomalous dimension), we have
\be
O(v)=e(v)=\frac{ig^2}{x^{+}(v)}- \frac{ig^2}{x^{-}(v)} \, .
\ee
Again, in the large spin limit, one has (with $n\geq 1$):
\be
\left. \frac{d^n}{dv^n} e(v) \right |_{\pm b} = O\left (\frac{1}{b^{n+1}}\right ) \, ,
\ee
together with the estimate (\ref {Zpmb}) for the counting function. It follows that the non-linear terms
contained in the expression (\ref {sumobs}) for the energy are $O(1/b)=O(1/s)$.
Therefore, if we neglect $O\left ( (\ln s )^{-\infty} \right )$ terms, we are allowed to work with only the linear expression
\be
\gamma (g,s,L)=\sum _{k=1}^{s} e(u_k)=-\int _{-b}^{b} \frac {dv}{2\pi} e(v)
Z^{\prime}(v)-\sum _{h=1}^{L-2} e(u_h^{(i)}) + O\left ( (\ln s )^{-\infty} \right ) \, , \label {sumen}
\ee
which we find convenient to write in terms of the one loop and higher than one loop
densities:
\be
\gamma (g,s,L)=\sum _{k=1}^{s} e(u_k)=-\int _{-b}^{b} \frac {dv}{2\pi} e(v)
\sigma _H(v) -\int _{-b_0}^{b_0} \frac {dv}{2\pi} e(v)
\sigma _0(v)-\sum _{h=1}^{L-2} e(u_h^{(i)}) + O\left ( (\ln s )^{-\infty} \right ) \, .
\ee
Extending the domains of integration to the real axis does not give problems, since we are
neglecting $O\left ( (\ln s )^{-\infty} \right )$ terms: we get
\be
\gamma (g,s,L)=\sum _{k=1}^{s} e(u_k)=-\int _{-\infty}^{+\infty} \frac {dv}{2\pi} e(v)
\sigma _H(v) -\int _{-\infty}^{+\infty} \frac {dv}{2\pi} e(v)
\sigma _0^{(s)}(v)-\sum _{h=1}^{L-2} e(u_h^{(i)}) + O\left ( (\ln s )^{-\infty} \right ) \, .
\label {andime}
\ee
Passing to Fourier transforms we have
\be
\hat e(k) = 2 g \sqrt{2}\pi  \frac{e^{-\frac{|k|}{2}}}{k} J_{1}(\sqrt{2}gk)
\ee
and, consequently,
\be
\gamma (g,s,L)= - \int _{-\infty}^{+\infty} \frac{dk}{4\pi ^2}2 g \sqrt{2}\pi  \frac{e^{-\frac{|k|}{2}}}{k} J_{1}(\sqrt{2}gk) \Bigl [ \hat \sigma _H(k) +\hat \sigma _0^{(s)}(k)
+ 2\pi \sum _{h=1}^{L-2}\cos{k u_h^{(i)}} \Bigr ] + O\left ( (\ln s )^{-\infty} \right ) \, .
\label {andimek}
\ee
Comparing (\ref {andimek}) with (\ref {Skeq}), we see that
\be
\gamma(g,s,L) = \frac{1}{\pi} \lim_{k \to 0} \hat \sigma _H(k) + O\left ( (\ln s )^{-\infty} \right ) \, ,  \label {kl}
\ee
which extends the Kotikov-Lipatov relation \cite {KL} to all the sublogarithmic $O\left ( (\ln s )^{-n} \right )$,
$n\geq 1$, contributions and allows to compute the high spin anomalous dimension
from the higher than one loop density.

For computational reasons, it is more convenient to use the function
\ba
S(k)&=&\frac {\sinh \frac {|k|}{2}}{\pi |k|} \Bigl \{ \hat \sigma _H(k)
- \pi  \frac {e^{-\frac {|k|}{2}}}{ \sinh \frac {|k|}{2}}  \sum _{h=1}^{L-2} \left[ \cos k u_h^{(i)}-\cos { k \bar u_h^{(i)}} \right]
\Bigr \} \, \Rightarrow \,\gamma(g,s,L) = 2 \lim_{k \to 0} S(k)  \, . \label{Sdef}
\ea
The function (\ref {Sdef}) satisfies the integral equation (for $k>0$)
\begin{eqnarray}
&&S(k)=\frac {L}{k}[1-J_0({\sqrt {2}}gk)]
- g^2 \int _{0 }^{+\infty}\frac {dt}{\pi }
e^{-\frac{t}{2}} \hat K (\sqrt{2}gk,\sqrt{2}gt) \cdot \nonumber \\
 &\cdot& \Bigl \{ \frac {\pi t}{\sinh \frac {t}{2}}S(t)-4\pi \ln 2 \ \delta (t)-\pi (L-2)
\frac {1-e^{\frac {t}{2}}}{\sinh \frac {t}{2}}-2\pi \frac {1-
e^{-\frac {t}{2}}\cos \frac {ts}{{\sqrt {2}}}}{\sinh \frac {t}{2}}
+ \label {Skeq2} \\
&+& \pi  \frac {e^{\frac {t}{2}}}{ \sinh \frac {t}{2}}  \sum _{h=1}^{L-2} \left[ \cos t u_h^{(i)} -1 \right] \nonumber
\Bigl \} = 4g^2 \ln s \hat K( \sqrt{2}gk, 0) + 4g^2 \int _{0 }^{+\infty} \frac{dt}{e^{t}-1} \hat K^{*} (\sqrt{2}gk,\sqrt{2}gt)+ \nonumber  \\
&+&  \frac {L}{k}[1-J_0({\sqrt {2}}gk)]+4g^2 \gamma _E \hat K( \sqrt{2}gk, 0) + g^2 (L-2) \int _{0 }^{+\infty}dt  e^{-\frac{t}{2}}\hat K (\sqrt{2}gk,\sqrt{2}gt)  \frac {1-e^{\frac {t}{2}}}{\sinh \frac {t}{2}}  - \nonumber \\
&-& g^2 \int _{0 }^{+\infty} {dt}
\hat K (\sqrt{2}gk,\sqrt{2}gt)
 \frac {\sum _{h=1}^{L-2} \left[ \cos t u_h^{(i)} -1 \right]}{ \sinh \frac {t}{2}}-
g^2 \int _{0 }^{+\infty}{dt}
e^{-\frac{t}{2}} \hat K (\sqrt{2}gk,\sqrt{2}gt)
  \frac { t}{\sinh \frac {t}{2}}S(t)
\, , \nonumber
\end{eqnarray}
where $\hat K^{*} (t,t')=\hat K(t,t')-\hat K(t,0)$. We remark that such equation depends on the position of the holes $u_h^{(i)}$ (but not on both $\bar u_h^{(i)}$, $u_h^{(i)}$) and that the quantity $u_h^{(i)}$ is also an unknown of the problem and has to be determined by solving equation (\ref {holcond}),
\be
Z(u_h^{(i)})=\pi (2h+1-L) \, , \quad h=1, \ldots , L-2 \, , \nonumber
\ee
which introduces non-linear effects in the equation for the density $S(k)$.
Equation (\ref {Skeq2}) is exact if we neglect - in the high spin limit - terms which are $O\left ( (\ln s )^{-\infty} \right )$ and is the main integral equation of this paper\footnote{Actually, for twist two there is a simple way to get also the $O\left (\frac{\ln s}{s}\right )$ and $O\left (\frac{1}{s}\right )$
terms from the first of (\ref {Skeq2}): it is sufficient to replace in the argument of the $\cos $ function the quantity $\frac{s}{\sqrt{2}}$, which represent the leading contribution to the
position of the external hole at large $s$, with the more accurate estimate \cite {BKM} $\frac{s}{\sqrt{2}} \left (1+\frac{\gamma +1}{2s} + O(1/s^2) \right )$. We then get the same result as (29) of \cite {FZ}.}. It will be our starting point in order
to investigate the high spin limit of (minimal) anomalous dimension of twist operators in the $sl(2)$ sector.

We will study two cases. First, we will consider the limit:
\be
s\rightarrow \infty \, , \quad L \ \ \ \textrm{fixed} \, . \label {Lfix}
\ee
Then, we will focus on the scaling limit (\ref {jlimit}) \cite {BGK}:
\be
s\rightarrow \infty \, , \quad L\rightarrow \infty \, , \quad j=\frac {L-2}{\ln s}  \, \quad {\mbox {fixed}} \, .
\nn
\ee
The case (\ref {Lfix}) will be reported in Section 4, the case (\ref {jlimit}) in Section 5.

\section{High spin at fixed twist}
\setcounter{equation}{0}

In this section we report our results on the fixed twist case (\ref {Lfix}).
We naturally make a move from equation (\ref {Skeq2}), in which the various contributions to the known ('forcing') term are separated according to their power of $\ln s$. As a consequence of the structure of the forcing term, the high spin expansion of the function $S(k)$ goes on as a series of logarithmic (inverse) powers (\ref {subl}):
\be
S(k)=\sum_{n=-1}^\infty S^{(n)}(k)  \, (\ln s)^{-n} + O\left ( (\ln s )^{-\infty} \right ) \, . \label {Ssubl}
\ee
Consequently, the anomalous dimension at high spin follows the same fate:
\ba
\gamma(g,s,L)= f(g) \ln s + f_{sl}(g,L) + \sum_{n=1}^\infty \gamma^{(n)}(g,L)  \, (\ln s)^{-n} + O\left ( (\ln s )^{-\infty} \right ) \, . \nonumber
\ea
Coming back to $S(k)$, the term proportional to $\ln s $ satisfies the BES linear integral equation,
\begin{eqnarray}
&&S^{(-1)}(k)=4g^2 \hat K( \sqrt{2}gk, 0) -
g^2 \int _{0 }^{+\infty}{dt}
e^{-\frac{t}{2}} \hat K (\sqrt{2}gk,\sqrt{2}gt)
  \frac { t}{\sinh \frac {t}{2}}S^{(-1)}(t)
\, , \label {BES}
\end{eqnarray}
whose solution determines the universal scaling function through $f(g)=2S^{(-1)}(0)$.
The BES equation was introduced in \cite {ES,BES} and thoroughly studied in the weak \cite {ES,BES} and the strong coupling \cite {BBKS, CK, BKK, KSV} limit. The interested reader can refer to these papers for a detailed
study of equation (\ref {BES}).

\medskip

The four subsequent terms - independent of $s$ - appear in the linear integral equation for the density which determines the virtual scaling function $f_{sl}(g,L)$. This equation was written in \cite {BFR} (equation 4.11), then it was re-obtained in \cite {FZ} and used there and in our contemporaneous paper \cite {FGR4} (where it appears as equation 3.3). In notations used in this paper it reads ($k\geq 0$)
\begin{eqnarray}
&&S^{(0)}(k)= 4g^2 \int _{0 }^{+\infty} \frac{dt}{e^{t}-1} \hat K^{*} (\sqrt{2}gk,\sqrt{2}gt)+ \nonumber  \\
&+&  \frac {L}{k}[1-J_0({\sqrt {2}}gk)]+4g^2 \gamma _E \hat K( \sqrt{2}gk, 0) + g^2 (L-2) \int _{0 }^{+\infty}dt  e^{-\frac{t}{2}}\hat K (\sqrt{2}gk,\sqrt{2}gt)  \frac {1-e^{\frac {t}{2}}}{\sinh \frac {t}{2}}  - \label {Skeqs0} \\
&-&
g^2 \int _{0 }^{+\infty}{dt}
e^{-\frac{t}{2}} \hat K (\sqrt{2}gk,\sqrt{2}gt)
  \frac { t}{\sinh \frac {t}{2}}S^{(0)}(t)
\, , \nonumber
\end{eqnarray}
the virtual scaling function $f_{sl}(g,L)$ being
\begin{equation}
f_{sl}(g,L)=2 S^{(0)}(0) \, . \label {E-s}
\end{equation}
As in the case of $f(g)$, an explicit expression for $f_{sl}(g,L)$, interpolating from weak to strong
coupling has not been found yet. What we did is expanding equation (\ref {Skeqs0}) in a systematic way for small $g$,
thus getting the weak coupling convergent
series for $f_{sl}(g,L)$ (formula (4.1) of \cite {FGR4}):
\begin{eqnarray}
&& f_{sl}(g,L) = (\gamma_E - (L-2) \ln 2)f(g)+
8 (2 L-7) \zeta (3) \, \left( \frac{g}{\sqrt 2} \right)^4  +
\nonumber  \\
&& -\frac{8}{3} \left(\pi ^2 \zeta (3) (L-4)+3 (21 L-62) \zeta (5)\right) \left( \frac{g}{\sqrt 2} \right)^6 + \nonumber \\
&& +\frac{8}{15} \left(\pi ^4 \zeta (3) (3 L-13)+75 (46 L-127) \zeta (7)+5 (11 L-32) \pi ^2 \zeta (5)\right) \left( \frac{g}{\sqrt 2} \right)^8 + \nonumber  \\
&& - \Bigg( \frac{128}{945} \pi ^6 \zeta (3) (11 L-49)+8 \left(2695 \zeta (9) L+16 \zeta (3)^3 L-7156 \zeta (9)+56 \zeta (3)^3\right) + \nonumber \\
&& +\frac{40}{3} (25 L-64)
   \pi ^2 \zeta (7)+\frac{8}{45} (103 L-310) \pi ^4 \zeta (5) \Bigg) \left( \frac{g}{\sqrt 2} \right)^{10} +   \nonumber  \\
&&+ \Bigg( \frac{32}{45} \pi ^4 \zeta (7) (295 L-772)+\frac{8}{3} \pi ^2 \left(1519 \zeta (9) L+24 \zeta (3)^3 L-3628 \zeta (9)-88 \zeta (3)^3\right) + \nonumber \\
&& +8 \left(33285 \zeta (11) L+536 \zeta (3)^2 \zeta (5) L-86082 \zeta (11)-1728 \zeta (3)^2 \zeta (5)\right) + \nonumber \\
&& +\frac{8}{945} (2023 L-6266) \pi ^6 \zeta
   (5)+\frac{8 (2956 L-13231) \pi ^8 \zeta (3)}{14175} \Bigg) \left( \frac{g}{\sqrt 2} \right)^{12}+\ldots .
\end{eqnarray}
In addition, in \cite {FGR4} we performed the strong coupling analysis, by means of analytical and numerical computations. For the first leading terms (formula (4.12) of \cite {FGR4}) we provided the following outcome:
\begin{equation}
f_{sl} (g, L) = 2 \sqrt{2} \,  g  \left[ \ln \frac{2 \sqrt 2 }{g} - c_1 -
\frac{3 \, \ln 2}{2 \sqrt{2} \pi \, g} \ln \frac{2 \sqrt 2 }{g}
+ \frac{c_0+(2-L)\pi}{2 \sqrt {2} \pi g}
 - \frac{\textrm{K}}{ 8 \pi^2 g^2} \ln \frac{2 \sqrt 2 }{g}+ \frac{k_{-1}}{2\sqrt{2}\,g^2}+ O(\frac{\ln g}{g^3}) \right]   \,\, ,
\label {Efinal}
\end{equation}
where $c_1=1$, $c_0=6\ln 2 -\pi$, $\textrm{K}=\beta(2)$ is the Catalan's constant and (see also \cite{FZ})
\begin{eqnarray}
k_{-1}= \frac{4 \textrm{K}-9 (\ln 2)^2}{4 \sqrt{2} \pi^2}= -0.0118253\dots \,\, .
\end{eqnarray}
Importantly, in this asymptotic expansion we could show that the only trace of the twist comes up in the piece $c_0+(2-L)\pi$ and thus cancels out completely, at order $O(s^0)$, in the
asymptotic (large $g$) expansion of  $\Delta -s=\gamma +L$. It follows that
the constant term (i.e. $O(g^0)$) in $\Delta -s=\gamma +L$ at order $O(s^0)$ is $\frac{ 6 \ln 2 + \pi}{\pi}$ for any
twist: this allowed successful comparisons with string theory results \cite{BFTT}, which does not
distinguish between null and small values of $L$.

An alternative analysis of the strong coupling limit can be found in \cite {FZ}. This computation was done adapting the method introduced in \cite {BKK} for $f(g)$.

\medskip

For what concerns $\gamma ^{(n)}(g,L)$, i.e. the $O((\ln s )^{-n})$, $n\geq 1$, contributions to the anomalous dimensions, they are 'driven' by the holes depending parts of (\ref {Skeq2}). As a consequence of
(\ref {kl}), one has
$\gamma ^{(n)}(g,L)=2S^{(n)}(0)$, where $S^{(n)}(k)$ satisfies the integral equation ($k>0$)
\ba
S^{(n)}(k)&=& - \left. g^2 \int _{0}^{+\infty} {dt}
\hat K (\sqrt{2}gk,\sqrt{2}gt)
 \frac {\sum _{h=1}^{L-2} \left[ \cos t u_h^{(i)} -1 \right]}{ \sinh \frac {t}{2}}
\right |_{(\ln s)^{-n}}
 - \nonumber \\
 &-& g^2 \int _{0 }^{+\infty}{dt}
e^{-\frac{t}{2}} \hat K (\sqrt{2}gk,\sqrt{2}gt)
  \frac { t}{\sinh \frac {t}{2}}S^{(n)}(t)
\, , \label {Snk}
\ea
with the symbol $|_{(\ln s)^{-n}}$ standing for the component proportional to $(\ln s)^{-n}$.
The analysis of this case was done in detail in the paper \cite {FGR5}. We report here the main results.

In order to fulfil condition (\ref {holcond}), the position of the internal holes has to expand in inverse powers of $\ln s$:
\be
u_h^{(i)} = \sum_{n=1}^\infty \alpha_{n,h} (\ln s)^{-n} + O\left ( (\ln s)^{-\infty} \right ) \, . \label {upos}
\ee
Introducing the (even) derivatives in zero of the (even) function $\sigma (u)=Z'(u)$, developing them in powers of $\ln s$,
\be
\frac{d^{2 q}}{d u^{2 q}} \sigma(u=0) = \sum_{n=-1}^\infty \sigma^{(n)}_{2q} \, (\ln s)^{-n} \, , \label {deriv}
\ee
and imposing the condition (\ref {holcond}) for the holes we eventually get
the following recursive equation for the unknowns $\alpha_{n,h}$,
\ba
\alpha_{p+1,h}  &  =  & -  \sum_{r=1}^{p} \frac{\sigma^{(-1)}_{r}}{ \sigma^{(-1)}_0} \sum_{\{ j_1, \dots, j_{p-r+1}\}}
\prod_{m=1}^{p-r+1} \frac{(\alpha_{m,h})^{j_m}}{j_m !} -\sum_{l=0}^{p-1}   \sum_{r=1}^{p-l} \frac{  \sigma^{(l)}_{r-1}}{ \sigma^{(-1)}_0} \sum_{\{ j_1, \dots, j_{p-r-l+1}\}}
\prod_{m=1}^{p-r-l+1} \frac{(\alpha_{m,h})^{j_m}}{j_m !}, \ \ \ p\geq 1 \nonumber \\
\alpha_{1,h}  &  =  & \frac{\pi (2h-1+L)}{\sigma^{(-1)}_0}, \ \ \ \ \ \ p=0 \, ,
\label{alphaiter}
\ea
where the $j_m$ contained in the second term of the right hand side are constrained by the conditions
$\sum\limits _{m=1}^{p-r+1}j_m=r+1$, $\sum\limits _{m=1}^{p-r+1}mj_m=p+1$ and the ones in the third term
by $\sum\limits _{m=1}^{p-r-l+1}j_m=r$, $\sum\limits _{m=1}^{p-r-l+1}mj_m=p-l$.

The next step is the Neumann expansion for the functions $S^{(n)}(k)$ (in the domain $k>0$):
 \be
S^{(n)}(k)=\sum _{p=1}^{\infty} S_p^{(n)}(g) \frac {J_p({\sqrt 2}gk)}{k} \, \Rightarrow
\gamma^{(n)}(g,L) = {\sqrt 2} g S_1^{(n)}(g) \, .
\ee
Straightforward but lengthy calculations, originating from equation (\ref {Skeq2}), lead to the conclusion
that the Neumann modes $S_p^{(n)}(g)$ satisfy the system\footnote {We use the notation:
\begin{equation}
Z_{n,m}(g)= \int _{0}^{+\infty}
\frac {dt}{t} \frac {J_{n}({\sqrt {2}}gt)J_{m}({\sqrt {2}}gt)}{e^t-1}
\, .
\end{equation}}
\begin{eqnarray}
S^{(n)}_{2p-1}(g)&=& -(2p-1) \int _{0}^{+\infty}
\frac{dt}{t} \frac {{\mathcal P}_n(g,t) \, J_{2p-1}({\sqrt {2}}gt)}{\sinh \frac {t}{2}} - 2(2p-1)
\sum _{m=1}^{\infty} Z_{2p-1,m}(g) S^{(n)}_m (g) \, , \nonumber \\
&& \label {Ssystem} \\
S^{(n)}_{2p}(g)&=& - 2p \int _{0}^{+\infty}
\frac{dt}{t} \frac { {\mathcal P}_n(g,t) \,   J_{2p}({\sqrt {2}}gt)}{\sinh \frac {t}{2}} -
4p\sum _{m=1}^{\infty} Z_{2p,m}(g) (-1)^m S^{(n)}_{m}(g) \, . \nn
\end{eqnarray}
In (\ref {Ssystem}) ${\mathcal P}_n (g,t)$ appears as a coefficient in the high spin expansion
\be
P(s,g,t)  =   \sum_{n=1}^\infty {\mathcal P}_n (g,t) (\ln s)^{-n} \,
\label{logexpan}
\ee
of the internal holes-depending function
\be
P(s,g,t)=\sum _{h=1}^{L-2} \left[ \cos t u_h^{(i)} -1 \right] \, , \label {Ps}
\ee
appearing in (\ref {Snk}). As a consequence of (\ref {alphaiter}),
${\mathcal P}_n (g,t)$ depends on the various coefficients $\alpha_{m,h}$ of (\ref {upos}) as
\be
{\mathcal P}_n (g,t) =
\sum_{r=1}^n  t^r \; \cos \frac{\pi r}{2}\sum_{\{ j_1, \dots, j_{n-r+1}\}}
\frac{ \sum\limits_{h=1}^{L-2}\prod_{m=1}^{n-r+1} (\alpha_{m,h})^{j_m}}{\prod_{m=1}^{n-r+1}  j_m !}, \ \ \ \ \
\sum _{m=1}^{n-r+1}j_m=r \, , \ \ \ \sum _{m=1}^{n-r+1}m j_m=n \, .
\ee
In order to study the system (\ref {Ssystem}), it is useful to introduce the
``reduced coefficients" $\tilde S^{(k)}_{p}$, defined in equations (4.23,4.24) of \cite{FGR3} as
solutions of the reduced systems
\begin{eqnarray}
\tilde S^{(k)}_{2p}(g)&=&{\mathbb I}_{2p}^{(k)}(g)-4p\sum _{m=1}^{\infty}Z_{2p,m}(g)(-1)^{m}\tilde S^{(k)}_{m}(g) \, ,  \nonumber \\
&& \label {redSeqn2} \\
\tilde S^{(k)}_{2p-1}(g)&=&{\mathbb I}_{2p-1}^{(k)}(g)- 2(2p-1)\sum
_{m=1}^{\infty}Z_{2p-1,m}(g)
\tilde S^{(k)}_{m}(g) \, , \nn
\end{eqnarray}
with the explicit (i.e. not depending on the densities $\sigma _{2q}^{(n)}$) forcing terms,
\begin{equation}
\label{intforterm}
{\mathbb I}_r^{(k)}= r \int _{0}^{+\infty} \frac {dh}{2\pi} h^{2k-1}
\frac {J_r ({\sqrt {2}}gh)}{\sinh \frac {h}{2}} \, .
\end{equation}
Indeed, solutions to the systems (\ref {Ssystem}) are linear combinations of the various
$\tilde S^{(k)}_{r}(g)$ with coefficients depending on $\alpha_{n,h}$. In particular, for what concerns
$\gamma^{(n)} (g)$, its expression in terms of $\tilde S^{(k)}_{1}$ and $\alpha_{m,h}$ reads
\be
\frac{\gamma^{(n)} (g)}{\sqrt2 \, g} = -2 \pi \sum_{r=1}^n  \tilde S^{(r/2)}_{1}\; \cos \frac{\pi r}{2}  \sum_{\{ j_1, \dots, j_{n-r+1}\}}
\frac{ \sum\limits_{h=1}^{L-2}\prod_{m=1}^{n-r+1} (\alpha_{m,h})^{j_m}}{\prod_{m=1}^{n-r+1}  j_m !} \, , \ \
\sum _{m=1}^{n-r+1}j_m=r \, , \ \  \sum _{m=1}^{n-r+1}m j_m=n \, . \label {gamma-n}
\ee
The various $\alpha_{n,h}$ are written in terms of the densities with the help of (\ref{alphaiter}), leaving
eventually $\gamma^{(n)} (g)$ as depending on $\tilde S^{(k)}_{1}$ and $\sigma^{(r)}_{2q}$, with $r\leq n-3$.
This last property is very important, since it makes possible to build up a recursive calculation scheme for
the $\gamma^{(n)} (g)$, opening the way to push the computation up to
the desired order in $\ln s$.

As an example in paper \cite {FGR5} we gave the following exact results for the first $\gamma^{(n)} (g)$:
\ba
\gamma^{(1)} (g,L) & = & 0 \, , \label {gamma1}\\
\gamma^{(2)} (g,L) & = & \sqrt 2 g \,  \frac{\pi^3}{3 \, (\sigma^{(-1)}_{0})^2} (L-3)(L-2)(L-1) \,  \tilde S^{(1)}_{1}(g) \, ,
\label {gamma2}\\
\gamma^{(3)} (g,L) & = & - 2 \sqrt 2 g \frac{\pi^3 \, \sigma^{(0)}_{0}}{3 \, (\sigma^{(-1)}_{0})^3} (L-3)(L-2)(L-1)  \, \tilde S^{(1)}_{1}(g) \, ,
\label {gamma3} \\
\gamma^{(4)} (g,L) & = & \sqrt 2 g \,  2 \pi  \Bigl \{ \Bigl [
 -\frac{\pi^2}{3 \, (\sigma^{(-1)}_{0})^2} \Bigl (  \frac{\sigma^{(1)}_{0} }{  \sigma^{(-1)}_{0}} -
 \frac{3}{2} \frac{(\sigma^{(0)}_{0})^2 }{  (\sigma^{(-1)}_{0})^2 }  \Bigl ) (L-3)(L-2)(L-1) -  \nonumber \\
&-&  \frac{\pi^4 \sigma^{(-1)}_{2} }{90 \, (\sigma^{(-1)}_{0})^5} (L-3)(L-2)(L-1)(5+3L(L-4))
 \Bigr ] \tilde S^{(1)}_1 \,  -  \nonumber \\
 &-& \frac{\pi^4}{360 \, (\sigma^{(-1)}_{0})^4} (L-3)(L-2)(L-1)(5+3L(L-4)) \, \tilde S^{(2)}_1  \Bigr \} \label {gamma4} \, .
\ea
Due to the aforementioned recursive properties,
such expressions can be explicitly computed in the weak and in the strong coupling limit.
Weak coupling expansions are presented in Appendix A of \cite {FGR5}.
The strong coupling leading term is given in Section 3 of that paper.

\section{High spin and large twist}
\setcounter{equation}{0}

We now want to study the anomalous dimension in the limit (\ref {jlimit}):
\be
s\rightarrow \infty \, , \quad L\rightarrow \infty \, , \quad j=\frac {L-2}{\ln s}  \, \quad {\mbox {fixed}} \, .
\nonumber
\ee
Since the number of internal holes becomes infinite, we need to treat the sum over them contained in (\ref {Skeq2})
by means of the technique discussed in Section 2. Before doing that,
we introduce two points $\pm c$ which separate the internal holes and the Bethe roots: since the number
of internal holes equals $L-2$, the 'separator' $c$ has to satisfy the following relation:
\be
Z(c)=\frac{1}{2}\int _{-c}^{c}dv \sigma (v)=-\pi j \, \ln s   \, . \label {norm}
\ee
Now we can specialise formula (\ref {sumexpr}) to the sum
\be
\sum _{h=1}^{L-2} \left[ \cos t u_h^{(i)} -1 \right] \, ,  \label {summa}
\ee
over the internal holes $u_h^{(i)} \in
[-c,c]$. Remember that in $[-c,c]$ no Bethe roots are present: therefore the right hand side of
(\ref {sumexpr}) is zero. We get
\ba
\sum _{h=1}^{L-2} \left[ \cos {t u_h^{(i)}} -1 \right]&=&-\int _{-c}^{c} \frac {dv}{2\pi}(\cos tv -1)
\sigma (v) - \frac {\pi}{6} \frac {t \sin tc}{\sigma (c)} - \nonumber \\
&-&  \frac {7\pi ^3}{360} \frac {t ^3 \sigma (c) \, \sin tc +3t^2 \sigma _1(c)\, \cos tc -3t \frac
{(\sigma _1(c))^2}{\sigma (c)}\sin tc +t \sigma _2 (c) \, \sin tc }{(\sigma (c))^4}
+ O\left ( \frac {j^n}{(\ln s )^5} \right )
= \nonumber \\
&=& -2 \int _{-\infty}^{+\infty} \frac {dk}{4\pi ^2} \hat \sigma (k) \left [ \frac {\sin (t+k)c}{t+k}
- \frac {\sin kc}{k} \right ]- \frac {\pi}{6} \frac {t \sin tc}{\sigma (c)} - \nonumber \\
&-& \frac {7\pi ^3}{360} \frac {t ^3 \sigma (c) \, \sin tc +3t^2 \sigma _1(c)\, \cos tc +t \sigma _2 (c) \, \sin tc }
{(\sigma (c))^4} + O\left ( \frac {j^3}{(\ln s )^3} \right )
\, , \label {sumev}
\ea
where $\sigma _m(c)$ denotes the $m$-th derivative of the density $\sigma (v)$ in $v=c$.
This expression has to be inserted in (\ref {Skeq2}) and worked out together with condition
(\ref {norm}): we get in general a non-linear integral equation for the quantity $S(k)$, related to the Fourier transform of the density
of Bethe roots and internal holes $\hat \sigma (k)$ through (\ref {Sdef}).

This joined analysis of (\ref {Skeq2}, \ref {norm}) gets simplified in the limit (\ref {jlimit}).
Indeed, it follows from the structure of (\ref {Skeq2}) that, in the case of limit (\ref {jlimit}), the function $S(k)$ expands as \cite {FRS,FIR,FGR5}
\be
S(k)=\sum _{r=-1}^{\infty} (\ln s )^{-r}\sum ^{\infty}_{n=0} S^{(r,n)}(k) j^n +
O\left ( (\ln s )^{-\infty} \right ) \label {Skj}
\ee
and, correspondingly, the anomalous dimension behaves as (\ref {gammaj}):
\be
\gamma (g,s,L)=\ln s \sum ^{\infty}_{n=0} f_n(g) j^n +
\sum _{r=0}^{\infty} (\ln s )^{-r}\sum ^{\infty}_{n=0} f_n^{(r)}(g) j^n +
O\left ( (\ln s )^{-\infty} \right ) \, . \nonumber
\ee
The functions $f_n(g)$, $f_n^{(r)}(g)$ are called generalised scaling functions. In particular
$f_0(g)$ coincides with the universal scaling function $f(g)$.

Similarly, the separator between internal holes and Bethe roots, $c$, enjoys the following scaling in the limit (\ref {jlimit}),
\be
c=\sum _{r=0}^{\infty} (\ln s )^{-r}\sum ^{\infty}_{n=1} c^{(r,n)} j^n \, . \label {cj}
\ee
The constants $c^{(r,n)}$ are connected to the various components $\sigma^{(r,n)}_{2q}$ in which the density and its derivatives in zero expand in the limit (\ref {jlimit}),
\be
\frac{d^{2 q}}{d u^{2 q}} \sigma(u=0) = \sum_{r=-1}^\infty \sum_{n=0}^\infty \sigma^{(r,n)}_{2q} \, (\ln s)^{-r} j^n \, , \label {sigmarn}
\ee
by means of (\ref {norm}):
\be
\frac{1}{2}\int _{-c}^{c}dv \sigma (v)=-\pi j \, \ln s  \nn \, .
\ee
For instance, using (\ref {cj}, \ref {sigmarn}) in (\ref {norm}), we get, for the first $c^{(r,n)}$:
\ba
&& c^{(0,1)}=-\frac {\pi} {\sigma ^{(-1,0)}} \, , \quad
c^{(0,2)}=\pi \frac {\sigma ^{(-1,1)}} {[\sigma ^{(-1,0)}]^2} \, , \quad
c^{(0,3)}=\frac {\pi ^3}{6} \frac {\sigma ^{(-1,0)}_2} {[\sigma ^{(-1,0)}]^4}-\pi \frac {[\sigma ^{(-1,1)}]^2} {[\sigma ^{(-1,0)}]^3} \, , \nn \\
&& c^{(1,1)}= \pi \frac { \sigma ^{(0,0)}}{[\sigma ^{(-1,0)}]^2} \, , \quad
c^{(1,2)}=-2\pi \frac { \sigma ^{(0,0)}\sigma ^{(-1,1)}}
{[\sigma ^{(-1,0)}]^3} \, , \nn \\
&& c^{(1,3)}= 3 \pi  \frac { \sigma ^{(0,0)}[\sigma ^{(-1,1)}]^2}
{[\sigma ^{(-1,0)}]^4}-\frac {2}{3} \pi ^3
\frac { \sigma ^{(0,0)}\sigma ^{(-1,0)}_2}
{[\sigma ^{(-1,0)}]^5}+\frac {\pi ^3}{6} \frac {\sigma ^{(0,0)}_2} {[\sigma ^{(-1,0)}]^4}\, , \nn \\
&& c^{(2,1)}=-\pi \frac {[\sigma ^{(0,0)}]^2} {[\sigma ^{(-1,0)}]^3} \, . \label {crn}
\end{eqnarray}
Therefore, once we take the limit (\ref {jlimit}) and use condition (\ref {norm}), we
easily get that equation (\ref {Skeq2}) splits in a set of equations,
one for every function $S^{(r,n)}(k)$. Remember that $S^{(-1,0)}(k)$ coincides with the BES density
(function $S^{(-1)}(k)$ of last section): therefore, it satisfies the BES linear equation (\ref {BES}), i.e.
\be
S^{(-1,0)}(k)= 4g^2 \hat K( \sqrt{2}gk, 0) -
g^2 \int _{0 }^{+\infty}{dt}
e^{-\frac{t}{2}} \hat K (\sqrt{2}gk,\sqrt{2}gt)
  \frac { t}{\sinh \frac {t}{2}}S^{(-1,0)}(t)
\, . \nonumber
\ee
Another particular case is the function $S^{(-1,1)}(k)$: this satisfies the linear equation
\begin{eqnarray}
&&S^{(-1,1)}(k)= \frac {1}{k}[1-J_0({\sqrt {2}}gk)]+g^2 \int _{0 }^{+\infty}dt  e^{-\frac{t}{2}}\hat K (\sqrt{2}gk,\sqrt{2}gt)  \frac {1-e^{\frac {t}{2}}}{\sinh \frac {t}{2}}  - \nonumber \\
&-&
g^2 \int _{0 }^{+\infty}{dt}
e^{-\frac{t}{2}} \hat K (\sqrt{2}gk,\sqrt{2}gt)
  \frac { t}{\sinh \frac {t}{2}}S^{(-1,1)}(t)
\, , \label {S11}
\end{eqnarray}
and determines the generalised scaling function $f_1(g)=2 S^{(-1,1)}(0)$.
Equation (\ref {S11}) was studied in paper \cite {FGR1}.

The last case that has to be treated separately is $r=j=0$. However, in this case, the density
is obtained from $S^{(0)}(k)$ ($(\ln s )^0$ contribution at fixed twist, see last section)
and $S^{(-1,1)}(k)$, by means of the equality
\be
S^{(0)}(k)=(L-2) S^{(-1,1)}(k)+S^{(0,0)}(k) \, .
\ee
In the three cases we have discussed up to now, the dynamics of the internal holes is not relevant:
for the required approximations the internal holes can be supposed all lying at the origin, i.e.
the sum (\ref {summa}) can be considered as vanishing.

The particular form of the expansion (\ref {cj}) with coefficients given by (\ref {crn}) is
however of fundamental importance for
all the remaining cases. They can be studied together, by means of the integral equation
\begin{eqnarray}
&&S^{(r,n)}(k)= - g^2 \int _{0 }^{+\infty} {dt}
\hat K (\sqrt{2}gk,\sqrt{2}gt)
\left. \frac {\sum _{h=1}^{L-2} \left[ \cos t u_h^{(i)} -1 \right]}{ \sinh \frac {t}{2}} \right |_{\frac{j^n}{(\ln s )^{r}} }- \nn \\
&-& g^2 \int _{0 }^{+\infty}{dt}
e^{-\frac{t}{2}} \hat K (\sqrt{2}gk,\sqrt{2}gt)
  \frac { t}{\sinh \frac {t}{2}}S^{(r,n)}(t)
\, , \label {Srn}
\end{eqnarray}
in which the sum over the internal holes is evaluated through (\ref {sumev}) and the
'separator' $c$ is expanded as in (\ref {cj}), with coefficients depending on the
quantities $\sigma ^{(r,n)}_{2q}$ and determined by the use of (\ref {norm}).

From the methodological point of view, we need to distinguish two cases.
The first one covers the values $r=-1, n \geq 2$ and $r=0, n \geq 1$.
In this case only the first term in the right hand side of (\ref {sumev}), which is linear in the density, is relevant. Therefore in the final equation non-linearity comes only from the non-linear dependence of $c$ on
the density and their derivatives in zero:
\begin{eqnarray}
&&S^{(r,n)}(k)=  g^2 \int _{0 }^{+\infty} {dt}
\left. \frac{\hat K (\sqrt{2}gk,\sqrt{2}gt)}{\sinh \frac {t}{2}}
\int _{-c}^{c} \frac {dv}{2\pi}(\cos tv -1)
\sigma (v) \right |_{\frac{j^n}{(\ln s )^{r}} }- \nn \\
&-& g^2 \int _{0 }^{+\infty}{dt}
e^{-\frac{t}{2}} \hat K (\sqrt{2}gk,\sqrt{2}gt)
  \frac { t}{\sinh \frac {t}{2}}S^{(r,n)}(t)
\, . \label {Srn1}
\end{eqnarray}
For instance, using (\ref {crn}) we get
for the first cases:
\ba
&& \int _{-c}^{c} \frac {dv}{2\pi}(\cos tv -1)
\sigma (v)\Bigl |_{\ln s \cdot j^2}=0 \, , \nn \\
&& \int _{-c}^{c} \frac {dv}{2\pi}(\cos tv -1)
\sigma (v)\Bigl |_{\ln s \cdot j^3}=\frac {1}{6}\pi ^2 \frac {t^2}{[\sigma^{ (-1,0)}]^2} \, , \quad
\int _{-c}^{c} \frac {dv}{2\pi}(\cos tv -1)
\sigma (v)\Bigl |_{\ln s \cdot j^4 } = -\frac {1}{3}\pi ^2
\frac {t^2 \sigma^{(-1,1)}}{[\sigma^{ (-1,0)}]^3} \, , \nn \\
&& \int _{-c}^{c} \frac {dv}{2\pi}(\cos tv -1)
\sigma (v) \Bigl |_{\ln s \cdot j^5}=  -\frac{\pi ^4 t^4}{120 [\sigma^{(-1,0)}]^4}+\frac{\pi ^2 [\sigma^{ (-1,1)}]^2 t^2}{2[\sigma^{ (-1,0)}]^4}-\frac{\pi ^4 \sigma^{ (-1,0)}_2 t^2}{30 [\sigma^{(-1,0)}]^5} \, , \nn \\
&& \int _{-c}^{c} \frac {dv}{2\pi}(\cos tv -1)
\sigma (v) \Bigl |_{\frac {j^r}{(\ln s)^0}}=0 \, , \quad r=1,2 \ \, , \quad \int _{-c}^{c} \frac {dv}{2\pi}(\cos tv -1)
\sigma (v) \Bigl |_{\frac {j^3}{(\ln s)^0}}=
-\frac {1}{3} \pi ^2  \frac { \sigma ^{(0,0)}}
{[\sigma ^{(-1,0)}]^3 } \tilde S_1^{(1)}(g) \, , \nn \\
&& \int _{-c}^{c} \frac {dv}{2\pi}(\cos tv -1)
\sigma (v) \Bigl |_{\frac {j^4}{(\ln s)^0}}=   \pi ^2   \frac { \sigma ^{(0,0)}\sigma
^{(-1,1)}} {[\sigma ^{(-1,0)}]^4 } \tilde
S_1^{(1)}(g) \, , \nonumber \\
&& \int _{-c}^{c} \frac {dv}{2\pi}(\cos tv -1)
\sigma (v) \Bigl |_{\frac {j^5}{(\ln s)^0}}=
- \Bigl [ \frac {1}{6} \Bigl ( \frac {\pi ^4}{5}
\frac {\sigma _2^{(0,0)}}{[\sigma ^{(-1,0)}]^5} +12 \pi ^2
\frac {\sigma ^{(0,0)}[\sigma ^{(-1,1)}]^2} {[\sigma
^{(-1,0)}]^5} -\pi ^4  \frac {\sigma ^{(0,0)} \sigma ^{(-1,0)}_2} {[\sigma ^{(-1,0)}]^6} \Bigr )
\tilde S_1^{(1)}(g) - \nonumber \\
&-& \frac {\pi ^4}{30}    \frac {\sigma ^{(0,0)}}{[\sigma
^{(-1,0)}]^5} \tilde S_1^{(2)}(g)\Bigr ] \, . \label {sumev1}
\ea
In the remaining cases, i.e. $r\geq 1$, the evaluation of the sum over the internal holes (\ref {summa})
involves also terms explicitly non-linear in the density. Thanks to formul{\ae} (\ref {sumexpr}, \ref {sumev}),
however, everything is under control and, for instance,
for the first values of $r,n$ we get
\ba
\sum _{h=1}^{L-2} \left[ \cos {t u_h^{(i)}} -1 \right]\Bigl |_{\frac {j^0}{(\ln s)^r}}&=& 0 \, , \quad r=1,2,3,4 \, ,
\nonumber \\
\sum _{h=1}^{L-2} \left[ \cos {t u_h^{(i)}} -1 \right]\Bigl |_{\frac {j}{\ln s}}&=& \frac {\pi ^2}{6} \frac {t^2}{(\sigma ^{(-1,0)})^2} \, , \quad
\sum _{h=1}^{L-2} \left[ \cos {t u_h^{(i)}} -1 \right] \Bigl |_{\frac {j^2}{\ln s}}= -\frac {\pi ^2}{3} t^2 \frac {\sigma ^{(-1,1)}}{(\sigma ^{(-1,0)})^3} \, , \nonumber \\
\sum _{h=1}^{L-2} \left[ \cos {t u_h^{(i)}} -1 \right]\Bigl |_{\frac {j^3}{\ln s}}&=& - \frac {\pi ^2}{2} t^2
\frac {(\sigma ^{(0,0)})^2}{(\sigma ^{(-1,0)})^4} -
\frac {\pi ^2}{6} \frac {t^2}{(\sigma ^{(-1,0)})^4}\left [ \frac {2}{3} \pi ^2 \frac {\sigma _2^{(-1,0)}}{\sigma ^{(-1,0)}}-3  (\sigma ^{(-1,1)})^2 + \frac {t^2}{6} \pi ^2 \right ] \, , \nonumber \\
\sum _{h=1}^{L-2} \left[ \cos {t u_h^{(i)}} -1 \right]\Bigl |_{\frac {j}{(\ln s)^2}}&=& -\frac {\pi ^2}{3} t^2\frac {\sigma ^{(0,0)}}{(\sigma ^{(-1,0)})^3} \, , \quad
\sum _{h=1}^{L-2} \left[ \cos {t u_h^{(i)}} -1 \right]\Bigl |_{\frac {j^2}{(\ln s)^2}}= \pi ^2 t^2 \frac {\sigma ^{(0,0)}\sigma ^{(-1,1)}}{(\sigma ^{(-1,0)})^4} \, , \nonumber \\
\sum _{h=1}^{L-2} \left[ \cos {t u_h^{(i)}} -1 \right]\Bigl |_{\frac {j}{(\ln s)^3}}&=& \frac {7 \pi ^4}{360}t^4 \frac {1}
{(\sigma ^{(-1,0)})^4}+ \frac {7 \pi ^4}{90}t^2 \frac {\sigma _2^{(-1,0)}}
{(\sigma ^{(-1,0)})^5}+ \frac {\pi ^2}{2} t^2 \frac {(\sigma ^{(0,0)} )^2}{(\sigma ^{(-1,0)})^4} \, .
\label {sumev2}
\ea
In both cases, next steps are the usual ones and details can be found in \cite {FGR3,FIR,FGR5}. First, we perform a Neumann expansion for the even functions $S^{(r,n)}(k)$, in the domain $k\geq 0$:
\be
S^{(r,n)}(k)=\sum _{p=1}^{\infty} S_p^{(r,n)}(g) \frac {J_p({\sqrt 2}gk)}{k} \, .
\ee
This implies that the generalised scaling functions are expressed as
\be
f_n^{(r)}(g)= {\sqrt 2} g S_1^{(r,n)}(g) \, .
\ee
The Neumann modes $S_p^{(r,n)}(g)$ satisfy linear systems and are linear combinations
of the ''reduced'' coefficients $\tilde S_p^{(k)}$ which are solutions of the systems (\ref {redSeqn2}).
The coefficients driving such linear combinations depend on the densities and their derivatives in zero,
$\sigma^{(r',n')}_{2q}$, with $r'\leq r , n'\leq n-1$. This property allows to find, step by step, exact expressions
for the generalised scaling functions $f_n^{(r)}(g)$ in terms of $\sigma^{(r',n')}_{2q}$ and $\tilde S_1^{(k)}$.
For the first of them we get the following results:
\ba
\frac{f_3 (g)}{\sqrt {2}g}&=&\frac {1}{3}\pi ^3 \frac {\tilde S_1^{(1)}(g)}{[\sigma^{ (-1,0)}]^2} \, , \quad
 \frac{f_4(g)}{\sqrt {2}g}= -\frac {2}{3}\pi ^3
\frac {\tilde S_1^{(1)}(g) \sigma^{(-1,1)}}{[\sigma^{ (-1,0)}]^3} \, , \nn \\
\frac{f_5 (g)}{\sqrt {2}g}&=& - \frac{\pi ^5 \tilde S_1^{(2)}(g)}{60 [\sigma^{(-1,0)}]^4}+\frac{\pi ^3 [\sigma^{ (-1,1)}]^2 \tilde S_1^{(1)}(g)}{[\sigma^{ (-1,0)}]^4}-\frac{\pi ^5 \sigma^{ (-1,0)}_2 \tilde S_1^{(1)}(g)}{15 [\sigma^{(-1,0)}]^5} \, , \nn \\
\frac{f_3^{(0)}(g)}{\sqrt {2}g}&=&-\frac {2}{3} \pi ^3  \frac { \sigma ^{(0,0)}}
{[\sigma ^{(-1,0)}]^3 } \tilde S_1^{(1)}(g) \, , \quad
\frac{f_4^{(0)}(g)}{\sqrt {2}g}=  2 \pi ^3   \frac { \sigma ^{(0,0)}\sigma
^{(-1,1)}} {[\sigma ^{(-1,0)}]^4 } \tilde
S_1^{(1)}(g) \, , \nonumber \\
\frac{f_5^{(0)}(g)}{\sqrt {2}g}&=& \Bigl [ -\frac {1}{3} \Bigl ( \frac {\pi ^5}{5}
\frac {\sigma _2^{(0,0)}}{[\sigma ^{(-1,0)}]^5} + 12 \pi ^3
\frac {\sigma ^{(0,0)}[\sigma ^{(-1,1)}]^2} {[\sigma
^{(-1,0)}]^5} -\pi ^5  \frac {\sigma ^{(0,0)} \sigma ^{(-1,0)}_2} {[\sigma ^{(-1,0)}]^6} \Bigr )
\tilde S_1^{(1)}(g) + \nonumber \\
&+& \frac {\pi ^5}{15}    \frac {\sigma ^{(0,0)}}{[\sigma
^{(-1,0)}]^5} \tilde S_1^{(2)}(g)\Bigr ] \, . \nn \\
f_0^{(r)} &=& 0 \, , \quad r=1,2,3,4 \, ; \quad
\frac {f_1^{(1)}} {{\sqrt {2}} g }=- \frac {\pi ^2}{3} \frac {\tilde S_1^{(1)}(g)}{(\sigma ^{(-1,0)})^2} \, , \quad
\frac {f_2^{(1)}} {{\sqrt {2}} g }= \frac {2\pi ^3}{3} \frac {\sigma ^{(-1,1)}}{(\sigma ^{(-1,0)})^3}\tilde S_1^{(1)}(g) \, , \nonumber \\
\frac {f_3^{(1)}} {{\sqrt {2}} g } &=& \frac {\pi ^3}{(\sigma ^{(-1,0)})^4}\left [ \left ( \frac {2}{9}\pi ^2
\frac {\sigma _2^{(-1,0)}}{\sigma ^{(-1,0)}} - (\sigma ^{(-1,1)})^2
+ (\sigma ^{(0,0)})^2 \right ) \tilde S_1^{(1)}(g) + \frac {\pi ^2}{18}\tilde S_1^{(2)}(g) \right ] \, ,\nn \\
\frac {f_1^{(2)}} {{\sqrt {2}} g } &=& \frac {2\pi ^3}{3} \frac {\sigma ^{(0,0)}}{(\sigma ^{(-1,0)})^3}\tilde S_1^{(1)}(g) \, , \quad
\frac {f_2^{(2)} }{{\sqrt {2}} g } =- 2\pi ^3  \frac {\sigma ^{(0,0)} \sigma ^{(-1,1)}}{(\sigma ^{(-1,0)})^4}\tilde S_1^{(1)}(g) \, , \nonumber \\
\frac {f_1^{(3)}} {{\sqrt {2}} g } &=&-  \frac {\pi ^3} {(\sigma ^{(-1,0)})^4}\left [ \frac {7 \pi ^2}{180} \tilde S_1^{(2)}(g) + \left ( \frac {7 \pi ^2}{45}  \frac {\sigma _2^{(-1,0)}}
{\sigma ^{(-1,0)}}+   (\sigma ^{(0,0)} )^2  \right ) \tilde S_1^{(1)}(g) \right ]
\, .\nonumber
\ea
Explicit expressions in the weak and strong coupling limit for the various $f_n^{(r)}(g)$ can be given without much ado, because of the iterative structure of the relations which determine them. They can be found in
\cite {FGR3,FIR,FGR5}.

In particular, strong coupling limit of the anomalous dimension is of interest since it can be checked against string theory calculations. For what concerns the function
\be
f(g,j)=\sum _{n=0}^{\infty}f_n(g)j^n \, ,
\ee
performing such a check is not a difficult task, after results by Alday and Maldacena \cite {AM}.
Introducing (at large $g$) the quantity
\be
m(g)=\frac {2^{\frac {5}{8}}\pi
}{\Gamma \left ( \frac {5}{4} \right )} g^{\frac {1}{4}}e^{-\frac
{\pi g}{\sqrt {2}}}\left [1+O\left (\frac{1}{g}\right )\right ] \, , \label {mgap}
\ee
in \cite {AM} it was proved that in the limit (\ref {jlimit}),
when $g\rightarrow \infty$, $j \ll g$, with
$j/m(g)$ fixed, the quantity $f(g,j)+j$ has to coincide with the energy density
of the ground state of the $O(6)$ non-linear sigma model with mass gap $m(g)$.
When $j/m(g) \ll 1$ we are
in the nonperturbative regime of the $O(6)$ non-linear sigma model. In this case the energy density
can be computed by using Bethe Ansatz related techniques. This computation has been systematically
performed in \cite {BF}. In order to have agreement between our calculations for $f(g,j)$
and computations of \cite {BF} for the $O(6)$ non-linear sigma model, we must have that the quantities
$\Omega_n(g)$ computed in that paper have to be related to $f_n(g)$ by the relation
$f_n(g)=2^{n-1}\Omega_n(g)$.
Our results \cite {FGR4} for the strong coupling limit of $f_3(g),f_4(g), f_5(g)$\footnote {In \cite
{FGR4} results for the strong coupling limit of $f_n(g)$ and checks with $O(6)$ non-linear sigma model results were
performed up to $n=8$.} are
\begin{eqnarray}
f_3(g)&=& \frac {\pi ^2}{24m(g)} + O\left ( e^{-\frac
{\pi g}{\sqrt {2}}} \right ) \, , \label {f3} \\
f_4(g)&=& -\frac {\pi ^2}{12 [m(g)]^2} {\cal S}_1 + O(1) \, ,  \label {f4} \\
f_5(g)&=&-\frac {\pi ^4}{640 [m(g)]^3}+\frac {\pi ^2}{8[m(g)]^3}[{\cal S}_1]^2 + O\left ( e^{\frac
{\pi g}{\sqrt {2}}} \right ) \, ,  \label {f5}
\end{eqnarray}
where we used the compact notations
\begin{equation}
{\cal S}_{2s+1}=\frac {1}{\pi ^{2s+1}}\sum _{n=0}^{\infty} (-1)^n
\left [ \frac {1}{\left (n+\frac {1}{2} \right )^{2s+1}}+ \frac
{1}{(n+1)^{2s+1}} \right ] \, , \label {calS}
\end{equation}
and agree with corresponding formul{\ae} contained in \cite {BF}.

\section{Conclusions}
\setcounter{equation}{0}

In this paper we have discussed our activity on the study of minimal anomalous dimension of twist operators in the $sl(2)$ sector of ${\cal N}=4$ SYM at high spin.
We preferred to give a general description of the methods we used: the reader can refer to the original papers for details and for a complete list of our results.

The main tool we used is integral equation (\ref {Skeq2}) for the density of Bethe roots.
This equation coincides with the exact non-linear integral equation equivalent to the ABA equations of the $sl(2)$ sector if we neglect terms going to zero faster than any inverse power of the logarithm of the spin.
Equation (\ref {Skeq2}) is linear, apart from the non-linear dependence of the internal holes positions $u_{h}^{(i)}$
on the counting function $Z(u)$. These non-linear effects are taken into account by inverting relation
 (\ref {holcond}): in the high spin limit such inversion is feasible because of the recursive properties of the equations (\ref {alphaiter}) determining the various $u_{h}^{(i)}$. This allows to give explicit expressions
 (\ref {gamma-n}) for the coefficients of the high spin expansion of the anomalous dimension.

When the twist goes to infinity, the internal holes are described by their density and contribute to
integral equation (\ref {Skeq2}) introducing explicitly non-linear terms. These non-linear terms are evaluated (see Section 5) by using properties and techniques of the non-linear integral equation, which are discussed in Section 2. Their contribution to main equation (\ref {Skeq2}) is obtained by applying formula (\ref {sumexpr}) and is reported in equations
(\ref {sumev1}, \ref {sumev2}). In this case also, recursive properties of the
equations determining the high spin limit of the density are crucial in order to obtain explicit expressions for the anomalous dimension.

\vspace {0.8cm}

{\bf Acknowledgements} We thank D.Bombardelli, G.Infusino, P.Grinza for
collaboration on the various papers reported here and N. Gromov for discussion on wrapping effects.
We acknowledge the INFN grant {\it Iniziative specifiche FI11} and {\it PI14}, the Italian University PRIN 2007JHLPEZ "Fisica Statistica dei Sistemi Fortemente Correlati all'Equilibrio e Fuori Equilibrio: Risultati Esatti e Metodi di Teoria dei Campi" for travel financial support.


\begin{thebibliography}{xx}







\bibitem{MWGKP}
J.M. Maldacena, {\sl The large N limit of superconformal field
theories and supergravity}, Adv. Theor. Math. Phys.
{\bf 2} (1998) 231 and hep-th/9711200 $\bullet $
S.S. Gubser, I.R. Klebanov, A.M. Polyakov, {\sl Gauge theory
correlators from non-critical string theory},
Phys.Lett. {\bf B428} (1998) 105 and hep-th/9802109 $\bullet $
E. Witten, {\sl Anti-de Sitter space and holography}, Adv. Theor.
Math. Phys. {\bf 2} (1998) 253 and hep-th/9802150;
\bibitem{GKPII}
S.S. Gubser, I.R. Klebanov and A.M. Polyakov,
  {\sl A semi-classical limit of the gauge/string correspondence},
  Nucl.Phys. {\bf B636} (2002) 99
  [arXiv:hep-th/0204051];

\bibitem{FT}
S.~Frolov and A.~A.~Tseytlin,
  {\sl Semiclassical quantization of rotating superstring in AdS(5) x S(5)},
  JHEP {\bf 0206} (2002) 007
  [arXiv:hep-th/0204226];





\bibitem{MZ}
J.A. Minahan, K. Zarembo, {\sl The Bethe Ansatz for ${\cal N}=4$
Super Yang-Mills}, JHEP{\bf 03} (2003) 013 and hep-th/0212208;

\bibitem{BS}
N. Beisert, M. Staudacher, {\sl The ${\cal N}=4$ SYM integrable super spin
chain}, Nucl. Phys. {\bf B670} (2003) 439 and hep-th/0307042 $\bullet $
V.A. Kazakov, A. Marshakov, J.A. Minahan, K. Zarembo,
{\sl Classical/quantum integrability in AdS/CFT},
JHEP{\bf 05} (2005) 024 and hep-th/0402207 $\bullet$
G. Arutyunov, S. Frolov, M. Staudacher, {\sl Bethe Ansatz for
quantum strings}, JHEP{\bf 10} (2004) 016 and hep-th/0406256 $\bullet $
M. Staudacher, {\sl The factorized S-matrix of CFT/AdS},
JHEP{\bf 05} (2005) 054 and hep-th/0412188 $\bullet $
N. Beisert, V.A. Kazakov, K. Sakai, K. Zarembo,
{\sl The Algebraic curve of classical superstrings on AdS(5) x S**5},
Commun. Math. Phys. {\bf 263} (2006) 659 and hep-th/0502226 $\bullet$
N. Beisert, M. Staudacher, {\sl Long-range $PSU(2,2|4)$ Bethe Ansatz
for gauge theory and strings}, Nucl. Phys. {\bf B727} (2005) 1 and
hep-th/0504190;


\bibitem{BES}
N. Beisert, B.Eden, M. Staudacher, {\sl Transcendentality and
crossing}, J.Stat.Mech.{\bf 07} (2007) P01021 and hep-th/0610251;

\bibitem{WRA}
J. Ambjorn, R. Janik, C. Kristjansen, {\sl Wrapping interactions and a new source of
corrections to the spin chain/string duality}, Nucl. Phys. {\bf B736} (2006) 288
and hep-th/0510171 $\bullet $
A.V. Kotikov, L.N. Lipatov, A. Rej, M. Staudacher, V.N. Velizhanin,
{\sl Dressing and wrapping}, J. Stat. Mech. {\bf 07} (2007) P10003 and
arXiv:0704.3586 [hep-th];

\bibitem{TBA}
D. Bombardelli, D. Fioravanti and R. Tateo,
 {\sl Thermodynamic Bethe Ansatz for planar AdS/CFT: a proposal},
  J.\ Phys.\ A  {\bf 42} (2009) 375401
and arXiv:0902.3930 [hep-th]
$\bullet$
N. Gromov, V. Kazakov, A. Kozak and P. Vieira,
{\sl Exact Spectrum of Anomalous Dimensions of Planar N = 4 Supersymmetric Yang-Mills Theory: TBA and excited states}, Lett. Math. Phys. {\bf 91} (2010) 265, cf. also
arXiv:0902.4458 [hep-th]
%
$\bullet$
G. Arutyunov and S. Frolov,
{\sl Thermodynamic Bethe Ansatz for the $AdS_5 \times  S^5$ Mirror Model},
JHEP{\bf 05} (2009) 068
and arXiv:0903.0141 [hep-th]
%
$\bullet$
D.~Bombardelli, D.~Fioravanti and R.~Tateo,
{\sl TBA and Y-system for planar $AdS_4/CFT_3$}, 
Nucl. Phys. {\bf B834} (2010) 543 and arXiv:0912.4715 [hep-th]
$\bullet$
N. Gromov and F. Levkovich-Maslyuk,
{\sl Y-system, TBA and Quasi-Classical strings in AdS(4) x CP3}, 
arXiv:0912.4911 [hep-th];

\bibitem{Y}
N. Gromov, V. Kazakov, P. Vieira,
{\sl Exact Spectrum of Anomalous Dimensions of Planar N=4 Supersymmetric Yang-Mills Theory},
Phys. Rev. Lett. {\bf 103} (2009) 131601 and arXiv:0901.3753 [hep-th];



\bibitem{BJL}
Z. Bajnok, R. Janik, T. Lukowski,
{\sl Four loop twist two, BFKL,
wrapping and strings},
Nucl. Phys. {\bf B816} (2009) 376 and
arXiv:0811.4448 [hep-th] $\bullet $
T. Lukowski, A. Rej, V.N. Velizhanin,
{\sl Five-Loop Anomalous Dimension of Twist-Two Operators},
Nucl. Phys. {\bf B831} (2010) 105 and
arXiv:0912.1624 [hep-th];





\bibitem{BGK}
A.V.Belitsky, A.S. Gorsky, G.P. Korchemsky, {\sl Logarithmic scaling
in gauge/string correspondence}, Nucl. Phys. {\bf B748} (2006) 24
and hep-th/0601112;

\bibitem {ES}
B. Eden, M. Staudacher, {\sl Integrability and transcendentality},
J. Stat. Mech. {\bf 11} (2006) P014 and hep-th/0603157;

\bibitem{FTT}
S. Frolov, A. Tirziu, A.A. Tseytlin,
{\sl Logarithmic corrections to higher
twist scaling at strong coupling from AdS/CFT},
Nucl. Phys. {\bf B766} (2007)
232 and hep-th/0611269;

\bibitem{AM}
L.F. Alday, J.M. Maldacena, {\sl Comments on operators with large
spin}, JHEP{\bf 11} (2007) 019 and arXiv:0708.0672 [hep-th];


\bibitem {K}
G.P. Korchemsky,
{\sl Asymptotics of the Altarelli-Parisi-Lipatov
Evolution Kernels of Parton Distributions},
Mod. Phys. Lett. {\bf A4}
(1989) 1257;


\bibitem{Poly}
A.M. Polyakov, {\sl Gauge Fields as Rings of Glue}, Nucl. Phys. {\bf B164} (1980) 171;

\bibitem {KM}
G.P. Korchemsky, G. Marchesini,
{\sl Partonic distributions for
large $x$ and renormalization of Wilson loops},
Nucl. Phys. {\bf
B406} (1993) 225 and hep-ph/9210281;



\bibitem{LIP}
L.N. Lipatov, {\sl Evolution equations in QCD}, in ``Perspectives in
Hadron Physics'', Prooceedings of the Conference, ICTP, Trieste, Italy,
May 1997, World Scientific (Singapore, 1998);
\bibitem{BDM}
V.M. Braun, S.E. Derkachov, A.N. Manashov, {\sl Integrability of three particle
evolution equations in QCD}, Phys. Rev. Lett. {\bf 81} (1998) 2020 and
hep-ph/9805225;



\bibitem{BBKS}
M. K. Benna, S. Benvenuti, I. R. Klebanov, A. Scardicchio, {\sl A
Test of the AdS/CFT Correspondence Using High-Spin Operators},
Phys. Rev. Lett. {\bf 98} (2007) 131603 and hep-th/0611135 $\bullet$
L. F. Alday, G. Arutyunov, M. K. Benna, B. Eden, I. R. Klebanov,
{\sl On the Strong Coupling Scaling Dimension of High Spin
Operators}, JHEP{\bf 04} (2007) 082 and hep-th/0702028 $\bullet $
I. Kostov, D. Serban and D. Volin, {\sl Strong coupling
limit of Bethe Ansatz equations},  Nucl.
Phys. {\bf B789} (2008) 413 and hep-th/0703031 $\bullet$
M. Beccaria, G.F. De Angelis, V. Forini, {\sl The scaling function
at strong coupling from the quantum string Bethe equations},
JHEP{\bf 04} (2007) 066 and hep-th/0703131;

\bibitem {CK}
P.Y. Casteill, C. Kristjansen, {\sl The strong coupling limit of
the scaling function from the quantum string Bethe Ansatz}, Nucl.
Phys. {\bf B785} (2007) 1 and arXiv:0705.0890 [hep-th];

\bibitem{BKK}
B. Basso, G.P. Korchemsky, J. Kotanski, {\sl Cusp anomalous
dimension in maximally supersymmetric Yang-Mills theory at strong
coupling}, Phys. Rev. Lett. {\bf 100} (2008) 091601 and
arXiv:0708.3933 [hep-th];
\bibitem{KSV}
I. Kostov, D. Serban and D. Volin, {\sl Functional BES equation},
JHEP{\bf 08} (2008) 101 and arXiv:0801.2542 [hep-th];

\bibitem{FRS}
L. Freyhult, A. Rej, M. Staudacher, {\sl A Generalized Scaling
Function for AdS/CFT}, J. Stat. Mech. (2008) P07015 and
arXiv:0712.2743 [hep-th];

\bibitem{BFR}
D. Bombardelli, D. Fioravanti, M. Rossi, {\sl Large spin
corrections in ${\cal N}=4$ SYM $sl(2)$: still a linear integral
equation}, Nucl. Phys. {\bf B810} (2009) 460 and arXiv:0802.0027 [hep-th];



\bibitem{FZ}
L. Freyhult, S. Zieme, {\sl The virtual scaling function of AdS/CFT},
Phys. Rev. {\bf D 79} (2009) 105009 and
arXiv:0901.2749 [hep-th];

\bibitem{FGR4}
D. Fioravanti, P. Grinza, M. Rossi, {\sl Beyond cusp anomalous
dimension from integrability}, Phys. Lett. {\bf B675} (2009) 137
and arXiv:0901.3161 [hep-th];

\bibitem{BFTT}
M. Beccaria, V. Forini, A. Tirziu, A.A. Tseytlin,
{\sl Structure of the large spin expansion of anomalous dimensions at
strong coupling}, Nucl. Phys. {\bf B812} (2009) 144 and
arXiv:0809.5234 [hep-th] $\bullet $ M. Beccaria, G.V. Dunne, V. Forini,
M. Pawellek, A.A. Tseytlin, {\sl Exact computation of one-loop correction to energy
of spinning folded string in $AdS_5 \times  S^5$}, J. Phys. A {\bf 43} (2010) 165402 and arXiv:1001.4018 [hep-th]
$\bullet $ S. Giombi, R. Ricci, R. Roiban, A.A. Tseytlin, C. Vergu,
 {\sl Generalised scaling function from light-cone gauge $AdS_5 \times  S^5$ superstring}, 
 arXiv:1002.0018 [hep-th];

\bibitem{GRO}
N. Gromov,
{\sl Generalized Scaling Function at Strong Coupling},
JHEP{\bf 11} 085 and arXiv:0805.4615 [hep-th];





\bibitem{FGR5}
D. Fioravanti, P. Grinza, M. Rossi, {\sl On the logarithmic powers of $sl(2)$ SYM$_4$},
Phys. Lett. {\bf B684} (2010) 52 and arXiv:0911.2425 [hep-th];



\bibitem{FGR1}
D. Fioravanti, P. Grinza, M. Rossi, {\sl Strong coupling for planar
 ${\cal N}=4$ SYM: an all-order result}, Nucl.
Phys. {\bf B810} (2009) 563 and arXiv:0804.2893 [hep-th];



\bibitem{BK}
B. Basso, G.P. Korchemsky, {\sl Embedding nonlinear $O(6)$ sigma
model into ${\cal N}=4$ super-Yang-Mills theory},
Nucl. Phys. {\bf B807}(2009) 397 and arXiv:0805.4194 [hep-th];

\bibitem{FGR2}
D. Fioravanti, P. Grinza, M. Rossi,
{\sl The generalised scaling function: a note},
Nucl. Phys. {\bf B827} (2010) 359 and arXiv:0805.4407 [hep-th];
\bibitem{FGR3}
D. Fioravanti, P. Grinza and M. Rossi,
{\sl The generalised scaling function: a systematic study},
JHEP{\bf 11} 2009 037 and arXiv:0808.1886 [hep-th];

\bibitem{FIR}
D. Fioravanti, G. Infusino and M. Rossi,
{\sl On the high spin expansion in the $sl(2)$ ${\cal N}=4$ SYM theory},
Nucl. Phys. {\bf B822} (2009) 467 and
arXiv:0901.3147 [hep-th];

\bibitem{AFS}
G. Arutyunov, S. Frolov, M. Staudacher, {\sl  Bethe ansatz for
quantum strings}, JHEP{\bf 10} (2004) 016 and hep-th/0406256;

\bibitem{BHL}
R. Janik, {\sl The $\text{AdS}_5\times\text{S}^5$ superstring
worldsheet S-matrix and crossing symmetry}, Phys. Rev. {\bf D73}
(2006) 086006 and hep-th/0603038 $\bullet $
N. Beisert, R. Hernandez, E. Lopez, {\sl A crossing symmetric phase
for $\text{AdS}_5\times\text{S}^5$ strings}, JHEP{\bf 11} (2006) 070
and hep-th/0609044;


\bibitem{BKP}
A.V. Belitsky, G.P. Korchemsky, R.S. Pasechnik,
{\sl Fine structure of anomalous dimensions in
${\cal N}=4$ super Yang-Mills theory},
Nucl. Phys. {\bf B809} (2009) 244 and
arXiv:0806.3657 [hep-ph];



\bibitem{FMQR}
D. Fioravanti, A. Mariottini, E. Quattrini, F. Ravanini, {\sl
Excited state Destri-de Vega equation for sine-Gordon and restricted
sine-Gordon models}, Phys. Lett. {\bf B390} (1997) 243
and hep-th/9608091
%
$\bullet $
G.~Feverati, F.~Ravanini and G.~Takacs,
  {\sl Truncated conformal space at c = 1, nonlinear integral equation and
  quantization rules for multi-soliton states},
  Phys.\ Lett.\  B {\bf 430} (1998) 264 and 
  hep-th/9803104 
%
 $\bullet $
 G.~Feverati, D.~Fioravanti, P.~Grinza and M.~Rossi,
 {\sl On the finite size corrections of anti-ferromagnetic anomalous  dimensions
  in N = 4 SYM}, JHEP {\bf 0605} (2006) 068 and 
  hep-th/0602189
 $\bullet $
%
G.~Feverati, D.~Fioravanti, P.~Grinza and M.~Rossi,
{\sl Hubbard's adventures in N = 4 SYM-land? Some non-perturbative
  considerations on finite length operators},
  J.\ Stat.\ Mech.\  {\bf 0702} (2007) P001
  and hep-th/0611186
%
$\bullet $
D.~Fioravanti and M.~Rossi,
  {\sl On the commuting charges for the highest dimension SU(2) operator in planar
  ${\cal N}=4$ SYM},
  JHEP {\bf 0708} (2007) 089 and
  arXiv:0706.3936 [hep-th];




\bibitem{KL}
A.V. Kotikov, L.N. Lipatov, {\sl On the highest transcendentality in
 ${\cal N}=4$ SUSY}, Nucl. Phys. {\bf B769} (2007) 217 and hep-th/0611204;

\bibitem{BKM}
A.V. Belitsky, G.P. Korchemsky, D. Mueller,
{\sl Towards Baxter equation in supersymmetric Yang-Mills theories},
Nucl. Phys. {\bf B768} (2007) 116 and  hep-th/0605291 $\bullet$
A.V. Belitsky,
{\sl Long-range SL(2) Baxter equation in N=4 super-Yang-Mills theory},
Phys. Lett. {\bf B643} (2006) 354 and
hep-th/0609068;





\bibitem{BF}
F. Buccheri, D. Fioravanti, {\sl The integrable $O(6)$ model and the
correspondence: checks and predictions}, arXiv:0805.4410 [hep-th].



\end{thebibliography}
\end{document}